\documentclass[osajnl,showpacs,superscriptaddress,10pt,amsmath,amssymb]{revtex4-1}

\usepackage{color}
\usepackage{graphicx}
\usepackage{citesort}

\newcommand{\comment}[1]{}

\newcommand{\ie}{\mbox{i.e.\ }}
\newcommand{\eg}{\mbox{e.g.\ }}

\newcommand{\eqnref}[1]{Eq.~(\ref{#1})}

\newcommand{\figref}[1]{Fig.~\ref{#1}}
\newcommand{\secref}[1]{Section~\ref{#1}}

\newcommand{\eps}{\varepsilon}
\newcommand{\Energy}{{\mathcal{E}}}
\newcommand{\Power}{{\mathcal{P}}}

\newcommand{\total}{{\rm{d}}}
\newcommand{\imag}{{\rm{i}}}

\newcommand{\Hamilton}{{\mathcal{H}}}
\newcommand{\perturb}{{\mathcal{V}}}

\newcommand{\PowerGain}{\Gamma}
\newcommand{\TotalGain}{\mathcal{A}_{\text{dB}}}

\newcommand{\cc}{{\rm{c.c.}}}

\newcommand{\myvec}[1]{{\bf{#1}}}

\newcommand{\Poisson}[2]{\Big\{ {#1} , \ {#2} \Big\}}

\newcommand{\SBS}{{\text{SBS}}}
\newcommand{\FSBS}{{\text{FSBS}}}
\newcommand{\BSBS}{{\text{BSBS}}}

\newcommand{\eff}{{\text{eff}}}
\newcommand{\tot}{{\text{tot}}}
\newcommand{\ac}{{\text{ac}}}
\newcommand{\opt}{{\text{opt}}}
\newcommand{\INT}{{\text{int}}}
\newcommand{\stat}{{\text{stat}}}
\newcommand{\dyn}{{\text{dyn}}}
\newcommand{\ideal}{{\text{ideal}}}
\newcommand{\direct}{{\text{dir}}}
\newcommand{\indir}{{\text{indir}}}
\newcommand{\bulk}{{\text{bulk}}}
\newcommand{\edge}{{\text{edge}}}

\begin{document}

\title{Brillouin resonance broadening due to structural variations in nanoscale waveguides}
\author{C. Wolff}
\affiliation{
  Centre for Ultrahigh bandwidth Devices for Optical Systems (CUDOS), 
}
\affiliation{
  School of Mathematical and Physical Sciences, University of Technology Sydney, NSW 2007, Australia
}

\author{R. Van Laer}
\affiliation{
  Photonics Research Group, Ghent University--imec, Belgium
}
\affiliation{
  Center for Nano- and Biophotonics, Ghent University, Belgium 
}

\author{M.~J. Steel}
\affiliation{
  Centre for Ultrahigh bandwidth Devices for Optical Systems (CUDOS), 
}
\affiliation{
  MQ Photonics Research Centre, Department of Physics and Astronomy, Macquarie University Sydney, NSW 2109, Australia
}

\author{B.~J. Eggleton}
\affiliation{
  Centre for Ultrahigh bandwidth Devices for Optical Systems (CUDOS),
}
\affiliation{
  Institute of Photonics and Optical Science (IPOS), School of Physics,
  University of Sydney, NSW 2006, Australia
}

\author{C.~G. Poulton}
\affiliation{
  Centre for Ultrahigh bandwidth Devices for Optical Systems (CUDOS),
}
\affiliation{
  School of Mathematical and Physical Sciences, University of Technology Sydney, NSW 2007, Australia
}

\email{christian.wolff@uts.edu.au}

\date{\today}

\begin{abstract}
  We study the impact of structural variations (that is slowly varying geometry 
  aberrations and internal strain fields) on the width and shape of the
  stimulated Brillouin scattering (SBS) resonance in nanoscale waveguides.
  We find that they lead to an inhomogeneous resonance broadening through two
  distinct mechanisms:
  firstly, the acoustic frequency is directly influenced via mechanical 
  nonlinearities;
  secondly, the optical wave numbers are influenced via the opto-mechanical 
  nonlinearity leading to an additional acoustic frequency shift via the
  phase-matching condition.
  We find that this second mechanism is proportional to the opto-mechanical 
  coupling and, hence, related to the SBS-gain itself.
  It is absent in intra-mode forward SBS, while it plays a significant role in 
  backward scattering.
  In backward SBS increasing the opto-acoustic overlap beyond a threshold 
  defined by the fabrication tolerances will therefore no longer yield the 
  expected quadratic increase in overall Stokes amplification.
  Finally, we illustrate in a numerical example that in backward SBS and 
  inter-mode forward SBS the existence of two broadening mechanisms with 
  opposite sign also opens the possibility to compensate the effect of 
  geometry-induced broadening.
  Our results can be transferred to other micro- and nano-structured waveguide
  geometries such as photonic crystal fibres.
\end{abstract}

\maketitle

%\ocis{
%  (190.5890) Nonlinear optics, scattering, stimulated;
%  (130.4310) Integrated optics, Nonlinear.
%}

%\bibliography{bibliography,extra}
%\bibliographystyle{osajnl}

%%%%%%%%%%%%%%%%%%%%%%%%%%%%%%%%%%%%%%%%%%%%%%%%%%%%%%%%%%%%%%%%%%%%%%%%%%%%%%%%
%%%%%%%%%%%%%%%%%%%%%%%%%%%%%%%%%%%%%%%%%%%%%%%%%%%%%%%%%%%%%%%%%%%%%%%%%%%%%%%%
\section{Introduction}

Stimulated Brillouin Scattering (SBS) is a nonlinear and self-amplifying 
interaction between guided optical waves and hypersound waves in 
waveguides~\cite{Boyd2003,Agrawal}.
The scattering of light from mechanical vibrations was first predicted by 
Brillouin in 1922~\cite{Brillouin1922}; the stimulated version of this process was first 
experimentally demonstrated~\cite{Chiao1964} shortly after the invention of the 
laser by Chiao et al. and has since been used successfully to characterize 
materials at hypersonic frequencies~\cite{Uchida1973}.
Following the general trend of miniaturization, SBS has been studied and
applied in ever smaller structures such as nano-structured optical 
fibres~\cite{Dainese2006} and even waveguides integrated on a 
chip~\cite{Eggleton2013}.
In these systems SBS ceases to be a bulk effect and surface effects strongly
come into play---most prominently radiation pressure appears as a second major
interaction process in addition to the bulk photoelastic 
effect.
Due to this additional coupling as well as the very tight mode confinement, 
high SBS-gains can be achieved in integrated 
waveguides~\cite{Pant2011,Rakich2012,Shin2013,vanLaer2015}, which makes them 
extremely interesting for signal processing applications~\cite{Chin2010} as 
well as narrow linewidth light sources~\cite{Abedin2012}.
Recently the idea of coherent phonon generation in phonon-lasers has gained 
considerable interest~\cite{Grudinin2010}.
However, all these potential applications require or at least greatly benefit 
from the extremely narrow linewidth of the SBS process; line width broadening 
is highly detrimental in many cases~\cite{Shin2013,vanLaer2015}.
This motivates our investigation of the inhomogeneous line broadening due to
structural variations of the waveguide and potential connections to the
SBS-gain itself.

The finite phonon life time 
(or equivalently in backward SBS: the phonon propagation length) 
defines an intrinsic line width of the SBS-resonance.
However, it has been known for some time that various imperfections of the 
SBS-system can distort the resonance and inhomogeneously broaden it.
Some of these imperfections such as insufficient acoustic guidance in 
combination with moderate optical mode confinement~\cite{Kovalev2002} 
(one possible explanation for SBS broadening in optical fibres, but somewhat 
debated in this context) or simply the finite length of suspended 
nanowires~\cite{Wolff2015b} are inherent to some waveguide designs and cannot 
be avoided.
In addition, local deviations from the intended ideal geometry can always lead
to local variations in the Stokes shift.
As a result, the total SBS-response of a non-ideal waveguide is the 
convolution of the intrinsic resonance and the Stokes frequency distribution
along the waveguide.
Incidentally, the geometry-dependence of the local Stokes shift has been used
to measure the homogeneity of optical fibres~\cite{Beugnot2011}
and to increase the SBS-threshold~\cite{Stiller2012}.
However, to date no connection has been made between the SBS-gain and the
sensitivity of the Stokes shift, although this is suggested by the close 
connection between SBS and opto-mechanics~\cite{vanLaer2015b}, in which in 
turn the coupling strength is known to be closely related to the 
deformation-sensitivity of the optical subsystem~\cite{Aspelmeyer2014}.

\begin{figure}
  \includegraphics[width=0.8\textwidth]{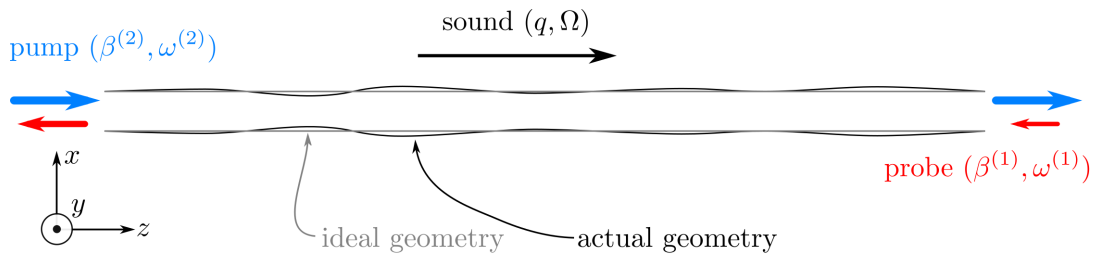}
  \caption{
    Sketch of the basic setup studied in this manuscript using the case of
    backward SBS and geometric variations of a simple rectangular waveguide
    as an example: 
    An optical pump (superscript $2$) and an optical probe (superscript $1$)
    excite a sound wave inside a waveguide that is assumed to be smooth, but
    whose geometry is smoothly and slowly perturbed along the axis of 
    propagation, \ie the $z$-axis.
    Other cases studied in this manuscript include forward SBS and 
    inhomogeneous strain fields (\eg due to temperature gradients or lattice
    mismatch) instead of the sketched geometry variations.
  }
  \label{fig:system}
\end{figure}

The aim of this paper is to understand the impact of smooth and slowly varying 
perturbations of the waveguide structure (as sketched in \figref{fig:system}).
We include variations of the waveguide geometry and internal strain fields that 
change smoothly along the waveguide and on a length scale greater than the
acoustic decay length.
This is opposed to surface roughness, which is usually a major concern in 
fabrication, since it causes linear optical propagation loss
via Rayleigh scattering---in backward SBS an analogous scattering of mechanical
waves also contributes to the acoustic loss.
In contrast, adiabatic structural variations do not scatter travelling waves
and do not cause linear loss.
However, they locally modify the dispersion relations and thereby influence 
the relative phase between the three waves participating in the SBS-process.
Here, we show that one mechanism for this is through the optical dispersion 
relation and the sensitivity is closely connected to the SBS-gain, whereas the
second mechanism is via the mechanical dispersion relation and completely 
independent from the SBS-gain.
We show that these two mechanisms influence forward and backward SBS quite 
differently and that increasing the SBS-gain either by increasing the 
acousto-optic coupling or reducing the mechanical damping leads to a 
regime where the line width is dominated by structural variations.
The total SBS-response is then reduced to the square root of the intrinsic 
SBS-gain.
We derive our results rigorously within the classical variant of a recent 
Hamiltonian formulation of SBS~\cite{Sipe2015}.

Throughout this manuscript we adhere to the notation and conventions 
introduced in our earlier paper on coupled-mode theory of SBS~\cite{Wolff2015a}.
In Section~\ref{sec:qualitative}, we qualitatively describe the relation between
SBS and the sensitivity of the dispersion relations and the resulting impact on
the SBS-resonance.
Although very useful for an intuitive understanding, this discussion lacks the
quantitative precision of the formal derivation provided in 
Section~\ref{sec:theory}.
The results of that section are then analytically studied in 
Section~\ref{sec:discussion}, focusing on the two important special 
cases of backward SBS and intra-mode forward SBS.
Next, we study the relative importance of the two broadening mechanisms in
forward and backward SBS in Section~\ref{sec:numerics} using a family of 
suspended silicon nanowires as an example and find that they can compensate each
other in appropriately engineered waveguides.
Section~\ref{sec:summary} summarises and concludes the paper.

%%%%%%%%%%%%%%%%%%%%%%%%%%%%%%%%%%%%%%%%%%%%%%%%%%%%%%%%%%%%%%%%%%%%%%%%%%%%%%%%
%%%%%%%%%%%%%%%%%%%%%%%%%%%%%%%%%%%%%%%%%%%%%%%%%%%%%%%%%%%%%%%%%%%%%%%%%%%%%%%%
\section{Qualitative description}
\label{sec:qualitative}

\begin{figure}
  \includegraphics[width=0.5\textwidth]{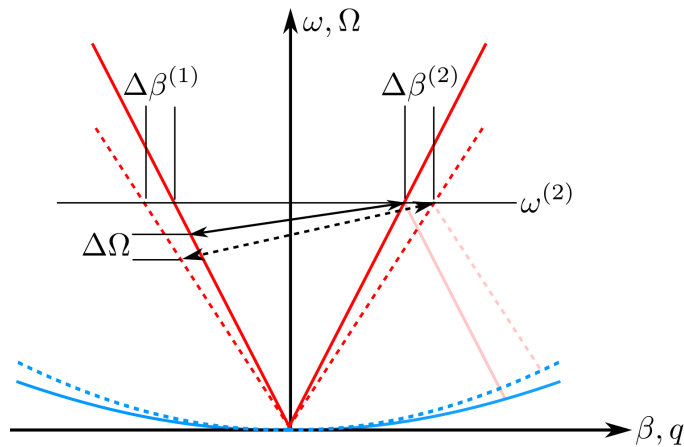}
  \caption{
    Sketch of how a perturbation of the optical and acoustic dispersion
    relations (red and medium blue, respectively) affects the Stokes shift
    via the phase matching condition for constant pump frequency $\omega^{(2)}$.
    the double-arrow highlights the opto-acoustic transition that fulfils
    the conservation of both energy and momentum.
    The solid lines represent the dispersion relations of the ideal system
    and the dashed lines those of the perturbed system.
    The light pink lines are parallels of the respective left side optical
    branches and indicate which acoustic wave number is phase matched.
    Note that the magnitude of $\Delta \Omega$ is grossly exaggerated compared
    to $\omega^{(2)}$.
  }
  \label{fig:phase_matching}
\end{figure}

The acoustic properties of a waveguide are modified by static structural 
perturbations via mechanical nonlinearities, which are usually not relevant 
for the dynamics of optically generated sound waves.
More precisely, there are separate contributions from internal strain fields 
and from variations of the cross section.
Strain fields enter via the nonlinear relation between stiffness and stress:
\begin{align}
  T_{ij} = \sum_{kl} c_{ijkl} S_{kl} + \sum_{klmn} f_{ijklmn} S_{kl} S_{mn},
\end{align}
where $S$ and $T$ are the strain and stress tensors, $c$ is the conventional
linear stiffness tensor and $f$ is a second order nonlinear stiffness tensor
(see \secref{sec:theory}).
At the same time, internal strain fields also modify the mass density 
distribution, because the trace of the strain tensor corresponds to a volume 
dilatation.
Finally, changes in the cross section geometry cause a change in the boundary
conditions (to be more specific: in the position of the boundary, not in the
type of condition imposed there) of the acoustic Helmholtz equation.
As an example, a slight reduction of the waveguide width might increase or 
decrease the acoustic frequency $\Omega$ depending on whether the
acoustic mode is predominantly longitudinal or transversal and on whether the
boundaries are approximately free or clamped.

At the same time, structural perturbations modify the optical properties of the
waveguide.
In contrast to the aforementioned acoustic frequency shift this modification 
is closely related to the SBS-process, which consists of two complementary 
nonlinear processes: the acoustic wave is excited by optical forces that are 
created by the beat of the optical pump and Stokes waves.
The sound wave spatio-temporally varies the waveguide's optical properties and 
modulates the pump wave such that the Stokes side band is amplified.
It is clear that the second process can only be efficient if the waveguide's
dispersion relation is sensitive to small mechanical perturbations; if it was
insensitive the acoustic wave would not modulate the pump wave.
This cannot be compensated by a stronger excitation of the sound wave 
(neglecting irreversible forces), because both processes are described by the 
same mode overlap integral, hence are equally strong. 
However, the optical dispersion relation cannot distinguish between static 
and dynamic (\ie acoustic) mechanical perturbations.
Therefore, the optical properties of any waveguide that exhibits strong SBS are
intrinsically sensitive to small structural perturbations introduced in the 
fabrication process.
More precisely, waveguides whose SBS is mainly caused by photoelasticity are
predominantly sensitive to internal strain fields, \eg caused by inhomogeneous
cooling of the sample, whereas waveguides whose SBS is mainly caused by 
radiation pressure are predominantly sensitive to slight variations of the
waveguide cross section.

The variation of the optical dispersion relation occurs as a function of $z$ 
rather than time, breaking the conservation of optical momentum.
Therefore, this variation results in a change in the optical wave numbers 
$\beta^{(1)}$ and $\beta^{(2)}$ while leaving the optical frequencies unchanged.
As a result, the pump frequency $\omega^{(2)}$ is exactly constant along the 
waveguide and the frequency $\omega^{(1)} = \omega^{(2)} - \Omega$ of the 
perfectly phase-matched Stokes wave varies only as a result of variations in 
the acoustic frequency $\Omega$.
These optical wave number variations translate to variations of the acoustic 
wave number through the momentum matching condition
\begin{align}
  q = \beta^{(2)} - \beta^{(1)}.
  \label{eqn:momentum_matching}
\end{align}
Therefore, the optical dispersion relation sensitivity indirectly affects the
acoustic frequency.
This effect has been practically applied in certain types of SBS-based fibre 
sensors~\cite{Horiguchi1995,Nikles1996,Beugnot2011}.
The total local change in the acoustic frequency 
(see also \figref{fig:phase_matching}) is given by the combination
of the optical wave number variations and the direct acoustic shift 
$\Delta \Omega^\direct$ described earlier:
\begin{align}
  \Delta \Omega^\tot = \Delta \Omega^\direct +
  \underbrace{
    \frac{\partial \Omega}{\partial q} (\Delta \beta^{(2)} - \Delta \beta^{(1)})
  }_{\Delta \Omega^\indir}.
  \label{eqn:tot_shift}
\end{align}
It should be noted at this point that the two contributions to the total 
frequency change can have opposite sign and therefore cancel (see 
Sec.~\ref{sec:numerics} for an example).

\begin{figure}
  \includegraphics[width=0.98\textwidth]{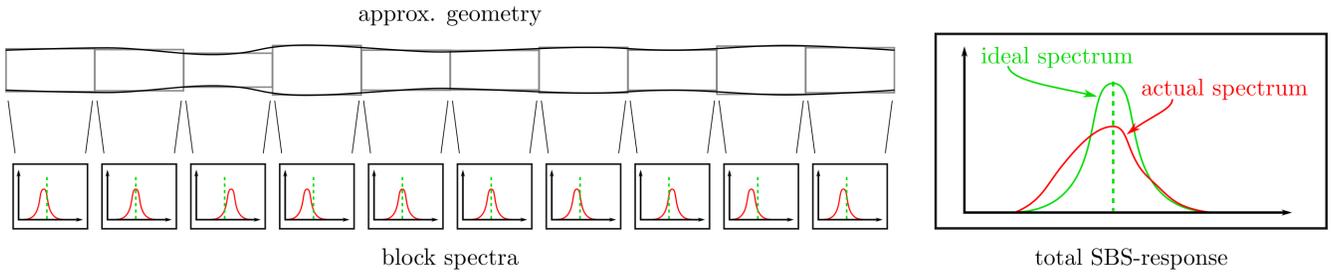}
  \caption{
    Sketch of how a spatial distribution of the acoustic resonance frequency
    causes an inhomogeneously broadened SBS-response of the total waveguide.
    The waveguide is approximated by a sequence of homogeneous waveguide 
    sections, each with its individual SBS-resonance centered around a frequency
    that might differ from the ideal frequency indicated by vertical dashed 
    lines.
    The total response (right panel) is the superposition of all block 
    responses and noticeably broader and shallower than the response of the 
    ideal structure if the acoustic frequency variance is comparable to the 
    resonance line width.
  }
  \label{fig:broadening}
\end{figure}

In a waveguide that is subject to inhomogeneous variations of the geometry
and strain fields the acoustic frequency variation becomes a function of the
position $z$ within the waveguide.
It is instructive to imagine the waveguide to be composed of many short
and approximately homogeneous sections (see \figref{fig:broadening}).
Each such section acts as an independent SBS-amplifier with a Stokes frequency
that is offset from the ideal by $\Delta \Omega^\tot(z)$.
All other properties such as intrinsic line width and differential SBS-gain
(\ie the SBS gain per unit length of waveguide measured in units of 
$\text{W}^{-1}\text{m}^{-1}$) can be assumed to be unchanged.
The total SBS-spectrum of the complete waveguide is then simply the 
superposition of the individual SBS-spectra.
In the limit of very short homogeneous sections, within an undepleted pump
approximation and assuming a Lorentzian intrinsic resonance shape, the Stokes 
amplification on resonance is then the integral:
\begin{align}
  \TotalGain(\Omega) = \int_0^L \total z \ 
  \PowerGain^\ideal\big(\Omega - \Delta \Omega^\tot(z)\big) P^{(2)}
  = \PowerGain^\ideal(\Omega) P^{(2)} \int_0^L \frac{\total z}{1 + 
  [2 \tau \Delta \Omega^\tot(z)]^2},
  \label{eqn:qualitative_superposition}
\end{align}
where $\tau$ represents the phonon life time and the waveguide is assumed 
to extend from $z=0$ to $z=L$.
Within the integral, $\PowerGain^\ideal$ is the local (differential) SBS
power gain measured in $\text{W}^{-1}\text{m}^{-1}$ (corresponding to the
quantity $g$ in Boyd's book~\cite{Boyd2003}) and $\TotalGain^\ideal(\Omega)$ is 
the total amplification accumulated over the total waveguide length $L$
(denoted $G$ by Boyd).
If the distribution $\Delta \Omega^\tot(z)$ has a width of the order 
$\tau^{-1}$ and 
especially if it is lopsided, the SBS-resonance is inhomogeneously broadened
into a non-Lorentzian shape as indicated in \figref{fig:broadening}.

So far, we have qualitatively explained the two main effects of structural 
variations on the SBS-resonance: perturbations to the acoustic and
the optical dispersion relation and we have furthermore argued that the 
magnitude of the latter effect must be closely related to the SBS-gain.
For practical purposes, it is probably sufficient to compute
$\Delta \Omega^\direct$ and $\Delta \beta^{(1,2)}$ numerically and insert them 
into Eqs.~(\ref{eqn:tot_shift},\ref{eqn:qualitative_superposition}).
However, the exact relationship between optical sensitivity and SBS-gain as 
well as further consequences can only be elucidated by a formal derivation.
This is therefore the topic of the next section.

\section{Quantitative derivation}
%%%%%%%%%%%%%%%%%%%%%%%%%%%%%%%%%%%%%%%%%%%%%%%%%%%%%%%%%%%%%%%%%%%%%%%%%%%%%%%%
\label{sec:theory}

We assume SBS in a lossless optical waveguide that is oriented along the 
$z$-direction and provides confinement for the optical and all relevant acoustic
modes, which are furthermore assumed to be only weakly damped.
Mostly out of convenience we restrict ourselves to mechanically isotropic
materials.
Furthermore, we assume that the optical and acoustic envelopes as well as the
deviations from the ideal waveguide geometry vary slowly along the $z$-axis
compared to the acoustic decay length.
We mostly follow the notation and conventions established in our earlier 
paper~\cite{Wolff2015a}:
upper case letters refer to physical (real-valued) quantities, lower case 
letters to eigenmodes and envelope functions, and a tilde implies that the
quantity is subjected to a phase transformation (rotating frame).
We furthermore introduce the symbol $\eta = \eps_r^{-1}$ for the inverse 
permittivity and the superscripts ``$\bulk$'' and ``$\edge$'' to refer to the 
interior of the waveguide's cross section and its boundary, respectively.
We use Einstein's summation convention where it helps to simplify the notation.

%%%%%%%%%%%%%%%%%%%%%%%%%%%%%%%%%%%%%%%%%%%%%%%%%%%%%%%%%%%%%%%%%%%%%%%%%%%%%%%%
\subsection{Hamiltonian}
\label{sec:hamiltonian}

We formulate this problem within a classical Hamiltonian framework.
To this end, we transfer a recent comprehensive quantum-mechanical description 
of SBS~\cite{Sipe2015} to the classical domain by replacing operators with 
functions and commutators with Poisson brackets.
One of the main results of Ref.~\cite{Sipe2015} is the full acousto-optic
Hamiltonian, which reads in classical form:
\begin{align}
  H = & \sum_{ijkl}\int \total^3 r \ 
  \Bigg[
    \frac{\Pi_i \Pi_i}{2\rho(\myvec r; \myvec U)}
    + \frac{1}{2} S_{ij} c_{ijkl}(\myvec r; \myvec U) S_{kl} 
  \Bigg]
  + \sum_{ij}\int \total^3 r \ 
  \Bigg[
    \frac{B_i B_i}{2 \mu_0} + 
    \frac{D_i \eta_{ij}(\myvec r; \myvec U) D_j}{2 \eps_0} 
  \Bigg]
    \quad = \int \total z \ \Hamilton(z) \ ,
  \label{eqn:initial_hamiltonian}
\end{align}
where $\Hamilton(z)$ is the Hamiltonian $z$-density, \ie the Hamiltonian per
unit length of waveguide.
The mechanical displacement, strain and momentum density fields and the 
electric and magnetic induction fields are denoted with $\myvec U$, 
$S_{ij}$, $\boldsymbol \Pi$, $\myvec D$ and $\myvec B$, respectively.
The material properties are described by the mass density $\rho$, the stiffness
tensor $c_{ijkl}$, the inverse relative permittivity tensor 
$\eta = \eps_r^{-1}$; the electromagnetic terms furthermore involve the
permittivity $\eps_0$ and permeability $\mu_0$ of vacuum.
All material constants are assumed to depend on the mechanical displacement 
field $\myvec U$ as well as its derivatives.
The first integral describes the mechanical part of the system, whereas the 
second integral covers both the electromagnetic part and the acousto-optic
interaction term, which is implicitly contained in the dependence of the 
inverse permittivity $\eta$ on the mechanical deformation field and its 
derivatives.
The two main contributions~\cite{Sipe2015} to this dependence are
\begin{align}
  \eta_{ij} \approx &
  \eta^\ideal_{ij}(\myvec r - \myvec U(\myvec r)) + \sum_{kl} p_{ijkl}(\myvec r)
  S_{kl}(\myvec r);
\end{align}
a further term of linear order in $\myvec U$ -- the moving polarization term 
described in Ref.~\cite{Wolff2015a} -- is usually too weak to be observable
and is neglected here.
Within this framework we describe structural perturbations as a static 
contribution to the total mechanical displacement field $\myvec U$:
\begin{align}
  \eta_{ij} \approx & 
  \eta^\ideal_{ij}(\myvec r - \myvec U^\stat - \myvec U^\dyn)
  + \sum_{kl} p_{ijkl} (S^\stat_{kl} + S^\dyn_{kl}),
  \label{eqn:ao_perturb}
\end{align}
where the superscripts ``$\stat$'' and ``$\dyn$'' denote the static and 
acoustic contributions, $\eta^\ideal$ is the permittivity distribution of 
the ideal (perfectly fabricated, sound-free) system and $p_{ijkl}$ is the
Pockels tensor that describes the photoelastic effect.
\eqnref{eqn:ao_perturb} reflects the close connection between the 
acousto-optic interaction and the sensitivity of the optical dispersion 
relation regarding structural perturbations.

Similarly, the mechanical stiffness and the mass density depend on the 
mechanical displacement field:
\begin{align}
  \rho(\myvec r) \approx & \rho^\ideal(\myvec r - \myvec U(\myvec r))
  - \sum_{i} \rho^\ideal(\myvec r) S_{ii}(\myvec r),
  \\
  c_{ijkl}(\myvec r) \approx & 
  c_{ijkl}^\ideal(\myvec r - \myvec U(\myvec r)) + 
  \sum_{mn} f_{ijklmn}(\myvec r) S_{mn}(\myvec r).
  \label{eqn:nonlinear_stiffness}
\end{align}
Here, the trace of the strain tensor expresses the volume dilatation.
The strain-dependent stiffness tensor presented in \eqref{eqn:nonlinear_stiffness} contains the leading order
nonlinearity of mechanically isotropic bulk materials; microscopically, 
the sixth rank tensor $f_{ijklmn}$ describes three-phonon scattering.
This can also be derived within the classical framework of 
hyperelasticity~\cite{Bigoni}:
in that approach, the stress is introduced as the strain-derivative of a free energy that 
is not quadratic in the strain variable.
Assuming a purely reversible mechanical response, this free energy is identical
to the Hamiltonian, whose strain-dependence can be expanded into a Taylor 
series.
The leading (cubic) anharmonic strain-term is created by the sixth rank tensor 
$f_{ijklmn}$ when inserting \eqnref{eqn:nonlinear_stiffness}
into \eqnref{eqn:initial_hamiltonian}.
In reality, this three-photon process might be forbidden by symmetry and a
higher-order multi-phonon scattering would constitute the leader order of
mechanical bulk nonlinearity. 
Furthermore, we have no reliable information about the order of magnitude of 
static strain fields across the large number of realisable integrated 
waveguides.
Still, we have introduced this three-phonon term to illustrate how mechanical bulk 
nonlinearities would impact inhomogeneous broadening, although we neglect them 
in our analysis in \secref{sec:discussion}.

Nonlinear mechanical wave processes usually can be neglected in typical SBS
setups, so we only retain the static contributions to the mechanical 
nonlinearity:
\begin{align}
  \rho(\myvec r) \approx & \rho^\ideal(\myvec r - \myvec U^\stat)
  - \sum_{i} \rho^\ideal(\myvec r) S^\stat_{ii},
  \\
  c_{ijkl}(\myvec r) \approx & 
  c_{ijkl}^\ideal(\myvec r - \myvec U^\stat) +
  \sum_{mn} f_{ijklmn}(\myvec r) S_{mn}^\stat.
\end{align}
These terms describe the sensitivity of the acoustic dispersion relation 
to structural perturbations.

We now decompose the total Hamiltonian $z$-density into the unperturbed 
mechanical and electromagnetic parts ($\Hamilton^\ac$, $\Hamilton^\opt$), 
the interaction term $\perturb^\INT$ and the structural perturbations 
$\perturb^\opt$ and $\perturb^\ac$:
\begin{align}
  \Hamilton = \Hamilton^\opt + \Hamilton^\ac + \perturb^\INT 
  + \perturb^\opt + \perturb^\ac,
\end{align}
where the first three terms are derived in Ref.~\cite{Sipe2015}:
\begin{align}
  \Hamilton^\ac = & \int \total^2 r \ \frac{\Pi_i \Pi_i}{2\rho^\ideal}
  + \frac{1}{2} S_{ij} c^\ideal_{ijkl} S_{kl} ,
  \\
  \Hamilton^\opt = & \int \total^2 r \ 
    \frac{B_i B_i}{2 \mu_0} + 
    \frac{D_i \eta^\ideal_{ij}(\myvec r) D_j}{2 \eps_0} ,
  \\
  \perturb^\INT = & \int \total^2 r \ \frac{D_i D_j}{2\eps_0} 
  \Big( p_{ijkl} S^\dyn_{kl} - U^\dyn_l \partial_l \eta^\ideal_{ij} \Big).\label{eqn:perturb_int}
\end{align}
The integrals are carried out over the entire transverse plane.  
The fourth term is formed in direct analogy to $\perturb^\INT$:
\begin{align}
  \perturb^\opt = & \int \total^2 r \ \frac{D_i D_j}{2\eps_0} 
  \Big( p_{ijkl} S^\stat_{kl} - U^\stat_l \partial_l \eta^\ideal_{ij} \Big),\label{eqn:perturb_opt}
\end{align}
and the last term is
\begin{align}
  \perturb^\ac = & \int \total^2 r \ \frac{S_{ij} S_{kl}}{2} \left(
  f_{ijklmn} S^\stat_{mn} - U_m^\stat \partial_m c_{ijkl}^\ideal
\right)
- \frac{\Pi_i \Pi_i}{2 (\rho^\ideal)^2}
\left( S_{jj}^\stat \rho^\ideal - U_j^\stat \partial_j \rho^\ideal \right).\label{eqn:perturb_ac}
\end{align}
The difference between $\perturb^\INT$ and $\perturb^\opt$ is that the former
results in a nonlinear coupling term of the optical and acoustic equations of
motion, whereas the latter causes a linear perturbation of the wave propagation 
within the optical fields.
The fifth term $\perturb^\ac$ does not have a counterpart, because we neglected
nonlinear wave effects within the propagating mechanical waves.

%%%%%%%%%%%%%%%%%%%%%%%%%%%%%%%%%%%%%%%%%%%%%%%%%%%%%%%%%%%%%%%%%%%%%%%%%%%%%%%%
\subsection{Modal expansion and approximation}

Next, we expand the electromagnetic and mechanical fields into the eigenmode
bases of the respective idealized waveguide problems~\cite{Wolff2015a,Sipe2015}
\begin{align}
  (\nabla_\perp + \imag \beta^{(i)} \hat z) \times (\nabla_\perp + \imag \beta^{(i)} \hat z)
  \times \tilde{\myvec e}^{(i)} =& \eps \mu_0 \big[\omega^{(i)}\big]^2 \tilde{\myvec e}^{(i)};
  \label{eqn:ewp_opt}
  \\
  (\nabla_\perp + \imag q \hat z)_j c_{ijkl} (\nabla_\perp + \imag q \hat z)_k
  \tilde u_l = & - \rho \Omega^2 \delta_{il} \tilde u_l.
  \label{eqn:ewp_ac}
\end{align}
In both equations the wave numbers along the waveguide serve as parameters, 
where the optical wave number $\beta^{(2)}$ has to be chosen such that $\omega^{(2)}$
matches the frequency of the modelled pump laser and the acoustic wave number 
is subsequently found through \eqnref{eqn:momentum_matching}.
In very extended systems such as conventional optical fibres, an infinite 
number of acoustic basis functions can contribute to the total acoustic SBS 
response.
In this case the treatment by Beugnot et al.~\cite{Beugnot2012} is a viable 
alternative to our description.
However, we here restricted ourselves to nano-scale waveguides, in which the
spectral separation between acoustic modes is greater than the SBS linewidth,
\ie in which only one or very few acoustic eigenmodes can be simultaneously 
excited by the two optical waves.
Consequently, we express the optical and acoustic fields as modulations of 
single basis functions~\cite{Wolff2015a} 
\begin{align}
  \myvec E(\myvec r, t) = &  
  \tilde{\myvec e}^{(1)} \tilde a^{(1)}(z,t) \exp(\imag \beta^{(1)} z) 
  + \tilde{\myvec e}^{(2)} \tilde a^{(2)}(z,t) \exp(\imag \beta^{(2)} z) + \cc,
  \label{eqn:expansion_opt}
  \\
  \myvec U^\dyn(\myvec r, t) = & 
  \tilde{\myvec u}(\myvec r, t) \tilde b(z,t) \exp(\imag q z) + \cc.
  \label{eqn:expansion_ac}
\end{align}
Here, $\tilde a^{(n)}$ and $\tilde b$ are the classical counterparts of the
quickly rotating envelope functions in Ref.~\cite{Sipe2015} and are related to 
the more common slowly varying envelope functions $a^{(n)}$ and $b$ via the 
relations
\begin{align}
  \tilde a^{(n)}(z,t) = & a^{(n)}(z,t) \exp(- \imag \omega^{(n)} t),
  \label{eqn:fast_a_def}
  \\
  \tilde b(z,t) = & b(z,t) \exp(- \imag \Omega t).
  \label{eqn:fast_b_def}
\end{align}
They fulfil the Poisson bracket relations
\begin{align}
  \Poisson{\tilde a^{(i)}(z,t)}{\tilde a^{(j)}(z',t')} = & 
  \Poisson{\tilde a^{(i)}(z,t)}{\tilde b(z',t')} = 
  \Poisson{\tilde a^{(i)}(z,t)}{\tilde b^\ast(z',t')} = 0,
  \\
  \Poisson{[\tilde a^{(i)}(z,t)]^\ast}{\tilde a^{(j)}(z',t')} = & 
  \frac{\imag \omega^{(i)}}{\Energy^{(i)}} \delta_{ij} \delta(z-z') \delta(t-t'),
  \\
  \Poisson{\tilde b^\ast(z,t)}{\tilde b(z',t')} = & 
  \frac{\imag \Omega}{\Energy_b} \delta(z-z') \delta(t-t').
\end{align}
The symbols $\Energy$ denote the respective modal energies per unit length
of waveguide~\cite{Wolff2015a}:
\begin{align}
  \Energy^{(i)} = & 2 \eps_0 \sum_{kl} 
  \int \total^2 r \ [\tilde e^{(i)}_k]^\ast \eps^\ideal_{kl} \tilde e^{(i)}_l,
  \\
  \Energy_b = & 2 \Omega^2 \int \total^2 r \ \rho^\ideal |\tilde{\myvec u}|^2.
\end{align}
Within the expansions (\ref{eqn:expansion_opt},\ref{eqn:expansion_ac}) the 
unperturbed contributions to the Hamiltonian $z$-density simply become
\begin{align}
  \Hamilton^\opt(z) + \Hamilton^\ac(z) 
  = & |\tilde a^{(1)}|^2 \Energy^{(1)} + |\tilde a^{(2)}|^2 \Energy^{(2)} 
  + |\tilde b|^2 \Energy_b .
\end{align}

Using the fact that the acoustic eigenmodes (indexed by superscript $n$) for a 
fixed wave number $q$ form a complete function set, we can expand the static
deformations:
\begin{align}
  \myvec U^\stat(\myvec r) = & 
  \sum_n \zeta^{(n)}(z) \tilde{\myvec u}^{(n)}(x,y) + \cc \ ;
  \label{eqn:perturb_expansion_U}
  \\
  S_{ij}^\stat(\myvec r) = & 
  \sum_n \xi^{(n)}(z) \tilde s_{ij}^{(n)}(x,y)
  + \cc \ ;
  \label{eqn:perturb_expansion_S}
  \\
  \text{with} \quad \tilde s_{ij}^{(n)} = & \frac{1}{2} \left[
    \partial_i \tilde u_j^{(n)} + \partial_j \tilde u_i^{(n)}
  \right].
\end{align}
In this we have introduced separate expansions for the distribution of a static
strain field and the deviation of the waveguide geometry, which contains both
effects of the static strain field and variations of the geometry due to
fabricational (\eg photolithography or etching) imperfections.
In analogy to Refs.~\cite{Wolff2015a,Sipe2015}, we now introduce separate modal overlap 
integrals to capture the effects of photoelasticity and of boundary 
displacement:
\begin{align}
  Q^{\bulk}_{ij;p} = & \eps_0 \int_\bulk \total^2 r \ \sum_{klmn}
  \eps_r^2 [e^{(i)}_k]^\ast e^{(j)}_l p_{klmn} \partial_m [u^{(p)}_n]^\ast ;
  \label{eqn:def_q_bulk}
  \\
  Q^{\edge}_{ij;p} = &
  \int_{\edge} \total \myvec r \ 
  \Big[
    (\eps_a - \eps_b) \eps_0 (\hat n \times \myvec e^{(i)})^\ast (\hat n \times \myvec e^{(j)})
    -
    (\eps^{-1}_a - \eps^{-1}_b) \eps^{-1}_0 (\hat n \cdot \myvec d^{(i)})^\ast (\hat n \cdot \myvec d^{(j)})
  \Big]
  (\hat n \cdot \myvec u^{(p)})^\ast .
  \label{eqn:def_q_edge}
\end{align}
Now, by substituting the modal expansions in (\ref{eqn:expansion_opt},\ref{eqn:expansion_ac}) 
and (\ref{eqn:perturb_expansion_U},\ref{eqn:perturb_expansion_S}) into the interaction
terms in (\ref{eqn:perturb_int},\ref{eqn:perturb_opt}), and integrating over the transverse plane, we find the interactions can be
reduced to the forms
\begin{align}
  \perturb^\INT(z) = & a^{(1)} [a^{(2)}]^\ast b \underbrace{
\Big( Q^{\bulk}_{12;n_b} + Q^{\edge}_{12;n_b} \Big)}_{Q_\SBS} + \cc
  \label{eqn:modal_interact}
  \\
  \perturb^\opt(z) = & \sum_n |a^{(1)}|^2 \Big( \zeta^{(n)} Q^{\bulk}_{11;n} + \xi^{(n)} Q^{\edge}_{11;n} \Big) 
  + |a^{(2)}|^2 \Big( \zeta^{(n)} Q^{\bulk}_{22;n} + \xi^{(n)} Q^{\edge}_{22;n} \Big),
  \label{eqn:modal_perturb_opt}
\end{align}
where $n_b$ is the index of the phase-matched acoustic mode.
The edge contribution $Q^\edge_{ij;n}$ to the perturbation term $\perturb^\opt$ 
is of course identical to the expression by Johnson~\cite{Johnson2002}.
Finally, we analogously express the acoustic structural perturbation term 
\begin{align}
  \perturb^\ac(z) = & |b|^2 \sum_n \Big( \zeta^{(n)} R^\bulk_n + \xi^{(n)} R^\edge_n \Big),
  \label{eqn:modal_perturb_ac}
\end{align}
in terms of a bulk overlap integral
\begin{align}
  R^\bulk_p = & \frac{1}{2} \int_\bulk \total^2 r \ \sum_{imn} 
  \Bigg\{ \sum_{jkl} 
  [\partial_i u^{(n_b)}_j]^\ast [\partial_k u^{(n_b)}_l] f_{ijklmn} 
  + \frac{\Omega^2 \big|u_i^{(n_b)}\big|^2}{2 \rho^\ideal}
  \Bigg\} \Big(\partial_m u^{(p)}_n + \partial_n e^{(p)}_m \Big),
\end{align}
which directly corresponds to the photoelastic perturbation overlap 
$Q^\bulk_{ij;n}$ and an edge integral $R^\edge_p$ that corresponds to the 
electromagnetic boundary displacement overlap $Q^\edge_{ij;n}$.
The explicit expression for $R^\edge_p$ is slightly involved and requires a 
short derivation; both can be found in Appendix~\ref{appx:mech_edge}.
Mathematically, this term contains the edge contributions from 
\eqnref{eqn:perturb_ac}, \ie the spatial derivatives of the stiffness and 
mass density functions.
Physically, it describes the simple fact that the transverse acoustic 
pattern---which is a standing wave in the transversal plane---is detuned 
very much like a drumhead upon size variations of the transverse acoustic
cavity.

Using the Poisson bracket relations, expanding the $z$-integral of the 
Hamiltonian density into the appropriate Taylor series~\cite{Sipe2015}
and  truncating the expansion to first order, we find the equations 
of motion:
\begin{align}
  \partial_t \tilde a^{(i)} = & \Poisson{\tilde a^{(i)}}{\Hamilton^\opt} + 
  \Poisson{\tilde a^{(i)}}{\perturb^\INT + \perturb^\opt}
  \\
  = & -\imag \omega^{(i)} \tilde a^{(i)} - 
  v^{(i)} \partial_z \tilde a^{(i)}
  + \Poisson{\tilde a^{(i)}}{\perturb^\INT + \perturb^\opt} ;
  \\
  \partial_t \tilde b = & \Poisson{\tilde b}{\Hamilton^\ac} + \Poisson{\tilde b}{\perturb^\INT + \perturb^\ac}
  \\
  = & -\imag \Omega \tilde b - v_b \partial_z \tilde b
  + \Poisson{\tilde b}{\perturb^\INT + \perturb^\ac} .
\end{align}

%%%%%%%%%%%%%%%%%%%%%%%%%%%%%%%%%%%%%%%%%%%%%%%%%%%%%%%%%%%%%%%%%%%%%%%%%%%%%%%%
\subsection{Inhomogeneous broadening in steady state}

We now have all tools to compute the spectrum of the steady state SBS response 
of an imperfect waveguide using the acoustic Green function~\cite{Wolff2015a}.
To this end, we assume a small signal approximation and a local acoustic 
response approximation, for which we assume that the modal envelopes 
$a^{(1)}$, $b$ as well as the perturbation envelopes $\zeta^{(n)}$, $\xi^{(n)}$ 
vary slowly compared to the acoustic decay length, which is of the order
$\alpha^{-1} \approx 50\,\mu\text{m}$ in backward SBS and is negligible in 
intra-mode forward SBS.
Following Ref.~\cite{Wolff2015a}, we introduce the inverse acoustic decay 
length $\alpha$ as well as an additional parameter $\kappa$ that describes a 
wave number offset between the ideally phase-matched Stokes mode and the probe 
wave that is injected in a seeded small-signal SBS-experiment.
The Poisson brackets involving the perturbation 
terms~(\ref{eqn:modal_perturb_opt},\ref{eqn:modal_perturb_ac}) are:
\begin{align}
  \Poisson{a^{(i)}}{\perturb^\opt} = & \imag v^{(i)} \Delta \beta^{(i)} a^{(i)},
  \\
  \Poisson{b}{\perturb^\ac} = & \imag v_b \Delta q^\direct,
\end{align}
expressed in terms of the structural detuning parameters
\begin{align}
  \Delta \beta^{(i)}(z) = & \frac{\omega^{(i)}}{\Power^{(i)}}
  \sum_n \Big( \zeta^{(n)} Q^\bulk_{ii;n} + \xi^{(n)} Q^\edge_{ii;n} \Big);
  \\
  \Delta q^\direct(z) = & \frac{\Omega}{\Power_b}
  \sum_n \Big( \zeta^{(n)} R^\bulk_{n} + \xi^{(n)} R^\edge_{n} \Big).
\end{align}
The steady state equations then follow:
\begin{align}
  \partial_z a^{(1)} - \imag (\Delta \beta^{(1)} + \kappa) a^{(1)} = & - 
  \frac{\imag \omega^{(1)} Q_\SBS}{\Power^{(1)}} a^{(2)} b^\ast
  \\
  \partial_z a^{(2)} - \imag \Delta \beta^{(2)} a^{(2)} = & - 
  \frac{\imag \omega^{(2)} Q_\SBS^\ast}{\Power^{(2)}} a^{(1)} b
  \\
  \partial_z b + (\alpha - \imag \Delta q^\direct) b
  = & - \frac{\imag \Omega Q_\SBS}{\Power_b} [a^{(1)}]^\ast a^{(2)},
\end{align}
with the SBS coupling parameter 
$Q_\SBS = Q^\bulk_{12;n_b} + Q^\edge_{12;n_b}$ from 
\eqnref{eqn:modal_interact} and the powers 
$\Power^{(i)} = v^{(i)} \Energy^{(i)}$ and $\Power_b = b_b \Energy_b$ of the
eigenmodes as introduced in Ref.~\cite{Wolff2015a}.
We then solve the acoustic equation using its Green function:
\begin{align}
  \nonumber
  b(z) = & - \frac{\imag \Omega Q_\SBS}{\Power_b} \int_{0}^{\infty} \total z' \ 
  \Big\{ [a^{(1)}(z-z')]^\ast 
  a^{(2)}(z-z') \exp\big(-[\alpha - \imag \Delta q^\direct(z-z')] z'\big) \Big\}.
  \label{eqn:b_convolution}
\end{align}
We assumed that the optical intensities as well as the structural envelopes 
vary slowly on the length scale $\alpha^{-1}$, so we can assume that the 
$\Delta q^\direct(z)$, $\Delta \beta^{(1)}(z)$, $\Delta \beta^{(2)}(z)$ and the
absolute values $\big|a^{(1)}\big|$ and $\big|a^{(2)}\big|$ are constant within
the convolution \eqref{eqn:b_convolution}; $\kappa$ is simply a constant.
We may therefore approximate the product of optical envelopes at the position 
$z-z'$ as that product at position $z$ modulo a phase factor that expresses the
relative beat $\Delta \beta^{(2)} - \Delta \beta^{(1)} - \kappa$:
\begin{align}
  [a^{(1)}(z-z')]^\ast a^{(2)}(z-z') 
  \approx &
  [a^{(1)}(z)]^\ast a^{(2)}(z) 
  \exp \big( \imag [\Delta \beta^{(2)}(z) - \Delta \beta^{(1)}(z) - \kappa] z' \big),
  \\
  \Rightarrow \quad
  b(z) \approx & - \frac{\imag \Omega Q_\SBS}{\Power_b} 
  [a^{(1)}(z)]^\ast a^{(2)}(z) L(z),
  \intertext{\centering with}
  L(z) = & \frac{1}{\alpha - \imag [\Delta q^\tot(z) - \kappa]}
  \\
  \Delta q^\tot = &
  \Delta q^\direct - \Delta \beta^{(1)} + \Delta \beta^{(2)}.
\end{align}
We can now insert this expression for the acoustic field into the optical 
equations.
Along the lines of Ref.~\cite{Wolff2015a}, we then transform them into a pair 
of equations for the optical powers $P^{(i)}(z) = \Power^{(i)} |a^{(i)}(z)|^2$: 
\begin{align}
  \partial_z P^{(1)} = & \PowerGain(z) P^{(1)} P^{(2)},
  \\
  \partial_z P^{(2)} = & -\PowerGain(z) P^{(1)} P^{(2)}.
\end{align}
The local SBS-gain function is given by
\begin{align}
  \PowerGain(z) = & 
  \underbrace{
    \frac{2 \omega \Omega |Q_\SBS|^2}{\Power^{(1)} \Power^{(2)} \Power_b \alpha}
  }_{\PowerGain^\ideal}
  \cdot
  \frac{\alpha^2}{\alpha^2 + [\Delta q^\tot(z) - \kappa]^2}.
  \label{eqn:local_gain}
\end{align}
Within a small signal approximation, \ie $\partial_z P^{(2)} \approx 0$, this
leads to a separable equation for the Stokes power $P^{(1)}$:
\begin{align}
  \frac{\partial_z P^{(1)}}{P^{(1)}} = & \PowerGain(z) P^{(2)};
  \\
  \Rightarrow \quad P^{(1)}(z) = & P^{(1)}(0) \exp\Big[ P^{(2)} 
  \int_0^z \total z' \ \PowerGain(z')\Big],
\end{align}
and an expression for the total Stokes amplification 
\mbox{
$\TotalGain = 10 \log_{10} (P^{(1)}_\text{out} / P^{(1)}_\text{in}) \ \text{dB}$
}
of the waveguide of length $L$:
\begin{equation}
  \boxed{
    \TotalGain = \frac{10}{\log 10}  P^{(2)} \PowerGain^\ideal 
    \int_0^L \total z' \ \frac{\alpha^2}{\alpha^2 + [\Delta q^\tot(z') - \kappa]^2}
  } 
  \label{eqn:theory_result}
\end{equation}
in units of dB.

In the simple case of $z$-independent aberration coefficients $\zeta^{(n)}$
and $\xi^{(n)}$ the integral in \eqnref{eqn:theory_result} becomes equal to one 
if the detuning $\kappa$ is chosen to be $\kappa = \Delta q^\tot$.
This means that to an excellent approximation the acoustic frequency and thereby
the SBS-resonance is simply shifted by the frequency
\begin{align}
  \Delta \Omega^\tot = v_b \Delta q^\tot = 
  \underbrace{v_b \Delta q^\direct}_{\Delta \Omega^\direct}
  + \underbrace{v_b (\Delta \beta^{(2)} - \Delta \beta^{(1)})}_{\Delta \Omega^\indir},
\end{align}
in agreement with \eqnref{eqn:tot_shift}.
In the case of inhomogeneous coefficients, the integral in 
\eqnref{eqn:theory_result} expresses the linear superposition of individual 
SBS-resonances and therefore the inhomogeneous resonance broadening in agreement 
with \eqnref{eqn:qualitative_superposition}.

%%%%%%%%%%%%%%%%%%%%%%%%%%%%%%%%%%%%%%%%%%%%%%%%%%%%%%%%%%%%%%%%%%%%%%%%%%%%%%%%
%%%%%%%%%%%%%%%%%%%%%%%%%%%%%%%%%%%%%%%%%%%%%%%%%%%%%%%%%%%%%%%%%%%%%%%%%%%%%%%%
\section{Special cases}
\label{sec:discussion}

We will now study the two practically most relevant special cases: intra-mode
forward SBS and backward SBS.
Qualitatively, inter-mode scattering behaves similarly to backward SBS.

%%%%%%%%%%%%%%%%%%%%%%%%%%%%%%%%%%%%%%%%%%%%%%%%%%%%%%%%%%%%%%%%%%%%%%%%%%%%%%%%
\subsection{Forward intra-mode SBS}
\label{sec:FSBS}

\noindent
First, we study the case of forward intra-mode SBS, \ie we assume
\begin{align}
  \myvec e^{(1)}(\vec r, t) \approx & \myvec e^{(2)}(\vec r, t);
  \\
  q \approx & 0; \quad v_b \approx 0.
\end{align}
In this case, the SBS coupling coefficient is
\begin{align}
  Q_\SBS = & Q^\bulk_{11;n_b} + Q^\edge_{11;n_b}.
  \label{eqn:fsbs_gain}
\end{align}
Furthermore, the optical wave number perturbations are identical:
\begin{align}
  \Delta \beta^{(2)} = \Delta \beta^{(1)} = & 
  \frac{\omega^{(1)}}{\Power^{(1)}} \sum_n 
  \Big( \zeta^{(n)} Q^\bulk_{11;n} + \xi^{(n)}_{11;n} \Big).
\end{align}
As a result, $\Delta\Omega^\indir = 0$, \ie the indirect resonance broadening 
via the optical dispersion relation is absent in intra-mode forward SBS.
Strictly speaking, this result is exact only if the optical group velocity is
identical at $\omega^{(1)}$ and $\omega^{(2)}$; in practice this is irrelevant
except for extremely dispersive optical modes, \eg in the slow light regime 
near band edges.
The remaining source of structural resonance broadening is the direct mechanical
contribution
\begin{align}
  \Delta \Omega^\direct = & v_b \frac{\Omega}{\Power_b} \sum_n 
  \Big(\zeta^{(n)} R^\bulk_{n} + \xi^{(n)} R^\edge_{n}\Big)
  \\
  = & \frac{\Omega}{\Energy_b} \sum_n 
  \Big(\zeta^{(n)} R^\bulk_{n} + \xi^{(n)} R^\edge_{n}\Big).
\end{align}
If we assume that internal strain fields are too weak to cause noticeable 
resonance broadening, we may assume
\begin{align}
  \Delta \Omega^\direct = & \frac{\Omega}{\Energy_b} \sum_n \xi^{(n)} R^\edge_{n},
\end{align}
\ie that the resonance detuning can be regarded as the simple geometric effect 
of changing the size of a transversal acoustic cavity.
The mechanical perturbation overlaps $R$ and the acousto-optic overlaps $Q$ are
in principle completely independent, so there is no fundamental reason why an
increase in the SBS-gain would increase the inhomogeneous broadening in the case
of intra-mode forward SBS.

%%%%%%%%%%%%%%%%%%%%%%%%%%%%%%%%%%%%%%%%%%%%%%%%%%%%%%%%%%%%%%%%%%%%%%%%%%%%%%%%
\subsection{General backward SBS}
\label{sec:BSBS}

This is fundamentally different in the case of backward SBS and (to  a lesser 
extent) in the case of general inter-mode SBS.
The optical modes that participate in conventional backward SBS are the 
counter-propagating partners of each other and therefore related via complex
conjugation:
\begin{align}
  \myvec e^{(1)}(\myvec r, t) = & [\myvec e^{(2)}(\myvec r, t)]^\ast;
  \quad
  \Power^{(1)} = - \Power^{(2)}.
  \label{eqn:modes_bsbs}
\end{align}
Therefore, the SBS coupling coefficient and the optical wave number 
perturbations are:
\begin{align}
  Q_\SBS = & Q^\bulk_{12;n_b} + Q^\edge_{12;n_b},
  \\
  \Delta \beta^{(2)} = - \Delta \beta^{(1)} = & \
  \frac{\omega^{(1)}}{\Power^{(1)}} \sum_n \Big( 
  \zeta^{(n)} Q^\bulk_{11;n} + \xi^{(n)} Q^\edge_{11;n} \Big),
\end{align}
while the expression for the direct acoustic detuning is identical to the
case of forward SBS.
The subscript indices of the individual overlap integrals 
$Q^\bulk_{ij;n}$ and $Q^\edge_{ij;n}$ are the mode labels of the involved 
two optical and the acoustic mode 
[see Eqs.~(\ref{eqn:def_q_bulk}--\ref{eqn:def_q_edge})].
It is in fact the acousto-optic forward-SBS coupling of the respective 
acoustic and optical modes, \ie basically the square root of the forwards
SBS gain for that combination of modes.
This time, the optical wave number perturbations add up and the total Stokes
shift becomes:
\begin{align}
  \Delta \Omega^\tot = & 2 \sum_n \Big[
  \zeta^{(n)} \Big( \frac{\Omega}{\Energy_b} R^\bulk_n + 
  \frac{v_b \omega^{(1)}}{v^{(1)} \Energy^{(1)}} Q^\bulk_{11;n} \Big)
  +
  \xi^{(n)} \Big( \frac{\Omega}{\Energy_b} R^\edge_n + 
  \frac{v_b \omega^{(1)}}{v^{(1)} \Energy^{(1)}} Q^\edge_{11;n} \Big)
  \Big].
\end{align}
The direct mechanical contribution is qualitatively identical to the previous
case.
Furthermore, we can assume that the predominantly longitudinal acoustic mode 
contributing to backward SBS is rather insensitive to slight geometry 
variations.
This suggests to focus on the indirect (optical) contribution to the broadening
and ignore the direct (acoustic) contribution.
As a result, the general result \eqnref{eqn:theory_result} then has the explicit
form (\ie $\kappa = 0$):
\begin{align}
  \TotalGain = & 
  \frac{20 P^{(2)} \omega^{(1)} \Omega \alpha}{\Power_b \log 10} 
  \int_0^L \total z \  \frac{
    \Big|Q^\bulk_{12;n_b} + Q^\edge_{12;n_b}\Big|^2
  }{
  \Big(\alpha \Power^{(1)} \Big)^2 +
  \Big( 2 \omega^{(1)} \sum_n \zeta^{(n)}(z) Q^\bulk_{11;n} + 
      \xi^{(n)}(z) Q^\edge_{11;n}
    \Big)^2
  }.
  \label{eqn:bsbs_general_result}
\end{align}
The mode coefficients $ Q^\bulk_{11;n} $ and $ Q^\edge_{11;n} $ in the 
denominator are mode overlaps that define the forward SBS gain and are
identical to the coefficients $Q^\bulk_{12;n}$ and $Q^\edge_{12;n}$ in the
numerator except for a complex conjugation of one of the optical modes 
courtesy of \eqnref{eqn:modes_bsbs}.

%%%%%%%%%%%%%%%%%%%%%%%%%%%%%%%%%%%%%%%%%%%%%%%%%%%%%%%%%%%%%%%%%%%%%%%%%%%%%%%%
\subsection{Backward SBS dominated by radiation pressure}
\label{sec:rad_BSBS}

%\TODO{Should we say something about weak broadening?}
The general BSBS result \eqnref{eqn:bsbs_general_result} is too convoluted to
interpret directly.
We therefore restrict ourselves to a very specific limiting case that 
nonetheless provides some insight into the general impact of geometrical 
variations in BSBS in nanoscale waveguides.
We assume that the acousto-optic interaction is dominated by the edge term
$Q^\edge_{11;n}$, \ie by radiation pressure.
We furthermore assume that the contributions $\xi^{(n_b)}$ of the phase matched
acoustic mode to the geometry perturbations provide a good estimate for the
total optical sensitivity $\Delta \beta^{(1)}$.
We finally assume that the geometry variations are completely random within a
certain interval.
Within this approximation $\xi^{(n_b)}(z)$ is uniformly distributed in an 
interval $[-\xi_0, \xi_0]$; other distributions of $\xi^{(n_b)}(z)$ clearly
lead to different broadened resonance shapes, but also exhibit the general 
trends described below.
These assumptions reduce \eqnref{eqn:bsbs_general_result} to the much simpler 
form
\begin{align}
  \TotalGain \approx & 
  \frac{5 P^{(2)} \Omega \alpha}{\Power_b \omega^{(1)} \log 10} 
  \Bigg| \frac{ Q^\edge_{12;n_b} }{Q^\edge_{11;n_b}} \Bigg|^2
  \frac{1}{2\xi_0} \int_{-\xi_0}^{\xi_0} \frac{\total \xi}{ \Theta^2 + \xi^2 }
  \\
  \text{with} \quad 
  \Theta = & \frac{\alpha \Power^{(1)}}{2 \omega^{(1)} Q^\edge_{11;n_b}}.
\end{align}
For the case of strong broadening ($\Theta \ll \xi_0$) the integral approaches 
the limit 
\begin{align}
  \frac{1}{2\xi} \int_{-\xi_0}^{\xi_0} \frac{\total \xi}{ \Theta^2 + \xi^2 }
  \ \longrightarrow \ \frac{\pi}{2 \xi_0 \Theta} = 
  \frac{\pi \omega^{(1)} Q^\edge_{11;n_b}}{\alpha \Power^{(1)} \xi_0},
\end{align}
and therefore the total amplification approaches the limit
\begin{align}
  \TotalGain \ \longrightarrow \  
  \underbrace{
    \frac{5 \pi \Omega}{\Power_b \Power^{(1)} \log 10} 
  }_{\text{constants}}
  \underbrace{
    \Bigg| \frac{ Q^\edge_{12;n_b} }{Q^\edge_{11;n_b}} \Bigg|^2
  }_{\text{gain ratio}}
  \underbrace{
    \frac{P^{(2)} Q^\edge_{11;n_b}}{\xi_0}
  }_{\text{scaling}}.
\end{align}
This result consists of a combination of eigenmode constants such as 
frequencies and mode powers, the ratio of the intrinsic forward and backward 
SBS gains and a term defining how the broadened SBS amplification scales with 
various design parameters, especially the fabrication tolerance $\xi_0$.
The gain ratio is always slightly below one, as we show in 
Appendix~\ref{appx:gain_ratio}, and we can simply assume it to be a constant.
The last term finally scales with the inverse of the acoustic damping, the 
fabrication tolerance and $Q^\edge_{11;n_b}$, which can be identified as being
proportional to the square root of the FSBS power gain using 
\eqnref{eqn:local_gain} and \eqnref{eqn:fsbs_gain}. 
Alternatively, we can also formulate the result as scaling with the square 
root of the BSBS power gain by using 
\begin{align}
  \Bigg| \frac{ Q^\edge_{12;n_b} }{Q^\edge_{11;n_b}} \Bigg|^2
  \frac{P^{(2)} Q^\edge_{11;n_b}}{\xi_0}
  = &
  \Bigg| \frac{ Q^\edge_{12;n_b} }{Q^\edge_{11;n_b}} \Bigg|
  \frac{P^{(2)} |Q^\edge_{12;n_b}|}{\xi_0}.
\end{align}
Both interpretations are basically equivalent, because FSBS and BSBS gain are 
closely related (Appendix~\ref{appx:gain_ratio}).

If we assume a waveguide with a significant SBS contribution due to radiation 
pressure and we assume the tolerance $\xi_0$ to be given, the reduction of 
$\Theta$ either by reducing the acoustic damping parameter $\alpha$ or by 
increasing the FSBS acousto-optic coupling $Q^\edge_{11;n_b}$ will lead into 
a regime where the BSBS amplification is defined by $Q^\edge_{11;n_b}$.
In practice, the acoustic loss can be reduced \eg by reducing the contact area 
of a nearly suspended waveguide to its substrate~\cite{vanLaer2015}; certain 
slot waveguide designs~\cite{vanLaer2014} on the other hand allow control of 
the radiation pressure coupling by varying the gap size.

%{\color{red} \bf [ Should we add plots illustrating the change of regime? 
%Should we provide examples that show the shape of the broadened spectra for
%various distributions of $\xi$?]}

%%%%%%%%%%%%%%%%%%%%%%%%%%%%%%%%%%%%%%%%%%%%%%%%%%%%%%%%%%%%%%%%%%%%%%%%%%%%%%%%
%%%%%%%%%%%%%%%%%%%%%%%%%%%%%%%%%%%%%%%%%%%%%%%%%%%%%%%%%%%%%%%%%%%%%%%%%%%%%%%%
\section{Numerical example}
\label{sec:numerics}

\begin{figure}
  \includegraphics[width=\textwidth]{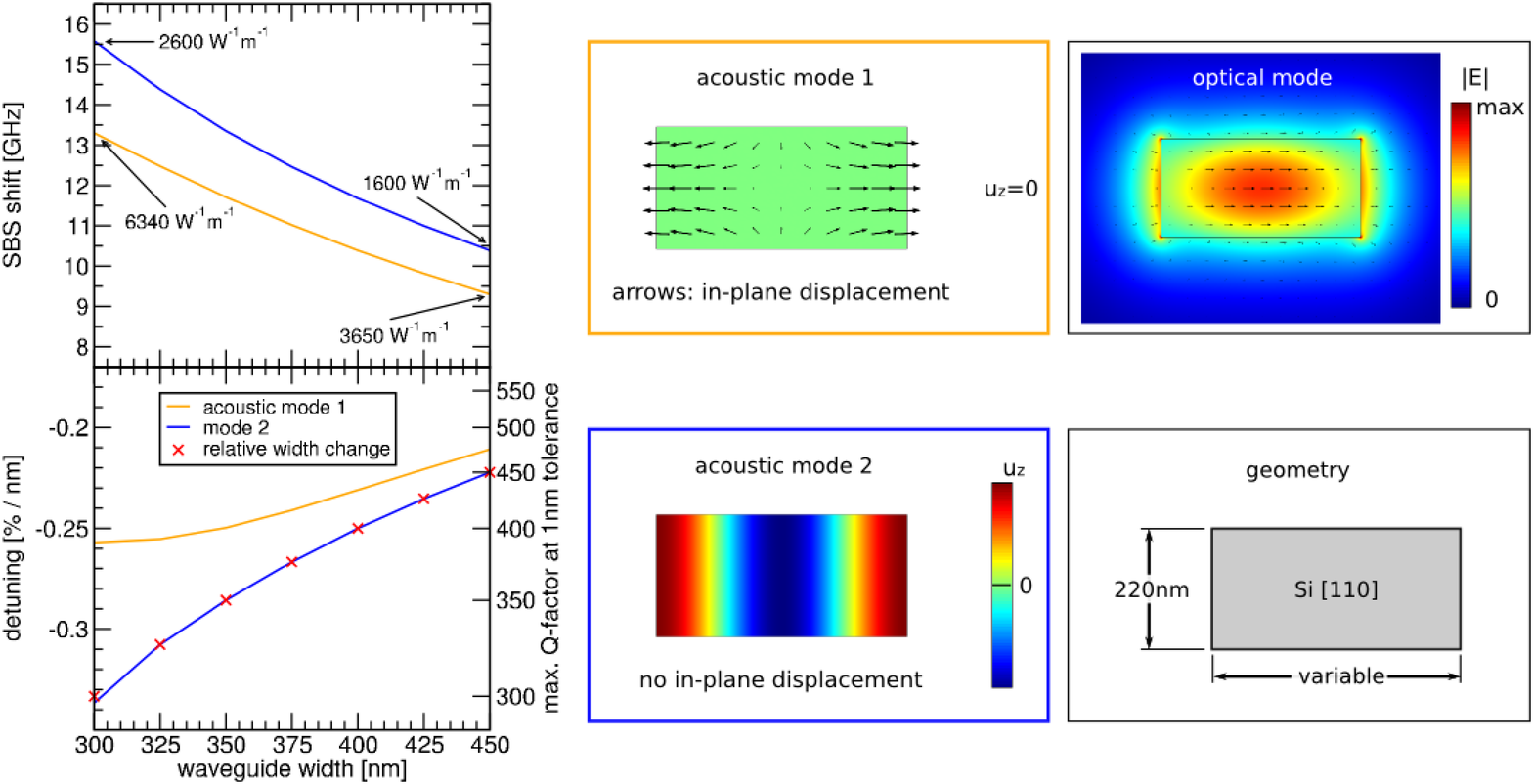}

  \caption{
    Numerical study of the sensitivity of \emph{forward} SBS (FSBS) to width 
    variations of a silicon nanowire oriented in the [110] crystal direction.
    We study the interaction between the fundamental optical mode and two 
    symmetry-permitted acoustic modes with lowest frequency. 
    The acoustic mode patterns show the in-plane displacement components as
    arrows and the axial displacement coordinate via a color map.
    The first acoustic mode (orange frame and graph lines) does not have any 
    axial contributions, whereas the second acoustic mode (blue frame and lines)
    has no in-plane contributions to the displacement field.
    The relative detunings are displayed in percent for a change of the waveguide
    width by $1\,\text{nm}$.
    Due to the extremely small acoustic wave number in FSBS only the direct 
    acoustic perturbation process contributes to the resonance shift, whose
    relative magnitude follows the relative width change exactly for mode 2 and
    qualitatively for mode 1.
    This demonstrates the direct acoustic response to width changes is 
    basically given by the detuning of the transverse mechanical cavity.
    The annotations in the top left panel specify the computed SBS power gains 
    of the two acoustic modes for the extreme waveguide widths assuming a
    mechanical quality factor of $300$.
  }
  \label{fig:example_fsbs}
\end{figure}

\begin{figure}
  \includegraphics[width=\textwidth]{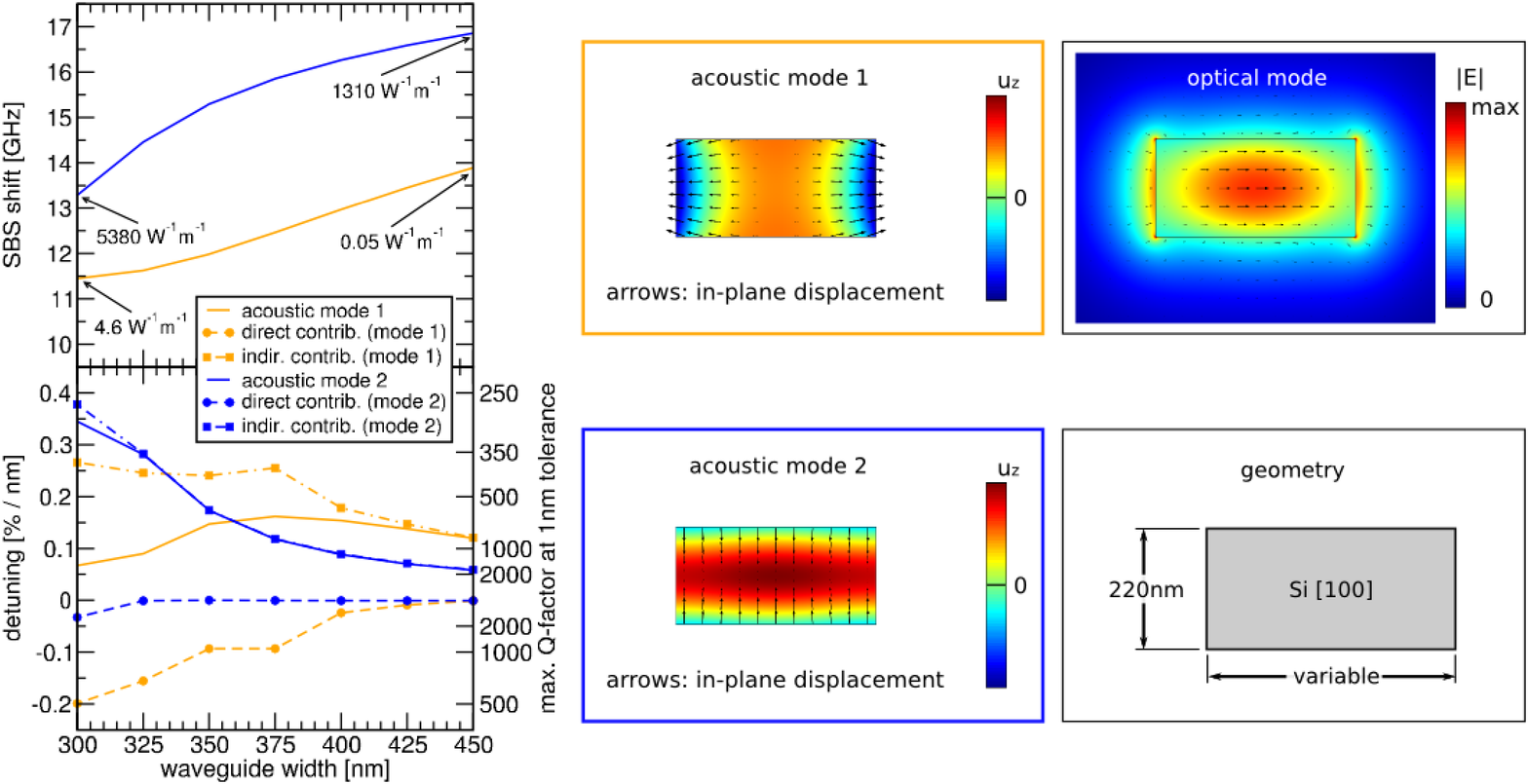}

  \caption{
    Numerical study of the sensitivity of \emph{backward} SBS (BSBS) to width
    variations of a silicon nanowire oriented in the [100] crystal direction.
    In contrast to FSBS (Fig.~\ref{fig:example_fsbs}), both studied acoustic 
    modes involve axial and in-plane components.
    Furthermore, BSBS features both the direct (acoustic) process (with negative 
    sign as for FSBS) and in addition the indirect (optical) process with the
    opposite sign.
    The former can be neglected for wider waveguides; however, it is possible
    that both contributions cancel each other.
    As in Fig.~\ref{fig:example_fsbs} we annotated the power gain, this time 
    assuming a quality factor of $1000$.
    The extremely low power gain of mode 1 is caused by nearly exact u
    cancellation of the photoelastic and moving-boundary related coupling in
    this structure.
  }
  \label{fig:example_bsbs}
\end{figure}

As the last part of our study, we performed numerical experiments on the 
sensitivity of the SBS frequency shift to variations of a waveguide's geometry.
We selected a family of silicon nanowires with a height of $220\,\text{nm}$ 
and a width varying between $300\,\text{nm}$ and $450\,\text{nm}$ in steps of 
$25\,\text{nm}$.
For each such waveguide, we characterized forward and backward SBS between the
fundamental optical mode and the two lowest symmetry-permitted acoustic 
modes~\cite{Wolff2014} using the finite-element solver COMSOL with an element 
size of $5\,\text{nm}$ inside the waveguide.
For each step of $25\,\text{mn}$, we estimated the width sensitivity by 
increasing the width by $1\,\text{nm}$ and repeating the calculation.
The resulting relative frequency shifts are plotted as solid lines in 
the lower left panels of Fig.~\ref{fig:example_fsbs} and 
Fig.~\ref{fig:example_bsbs}.
%These values agree with the derivative of $\Omega$ with respect to the waveguide
%width, as one would expect.
Finally, we added computed SBS power gains of the respective acoustic modes as
annotations to the SBS frequency graphs; they are intended to give a rough idea
of the respective acousto-optic coupling.

In the case of FSBS (Fig.~\ref{fig:example_fsbs}), the difference is entirely 
due to the direct (acoustic) perturbation effect, because the acoustic wave 
number is basically zero.
Both acoustic modes are standing waves in the acoustic resonator, formed 
perpendicular to the axis of light propagation, where the first mode has 
contributions from both in-plane components of the mechanical displacement 
field and the second mode only contributions from the out-of-plane component.
As a result, the second mode is a pure standing S-wave and consequently its
eigenfrequency is entirely determined by the waveguide width and the transverse
speed of sound.
This leads to the exact correspondence of the variation of $\Omega$ and the 
relative change of the waveguide width (red crosses in Fig.~\ref{fig:example_fsbs}).
In contrast, the first mode is a hybrid of an S-wave and a P-wave, because both
in-plane components contribute. 
As a result, the relationship between waveguide width and $\Omega$ is not as 
straightforward.
This mode for a waveguide width of $450\,\text{nm}$ was studied experimentally 
in Ref.~\cite{vanLaer2015} and its sensitivity has been determined to be 
$20\,\text{MHz}/\text{nm}$ at a measured Brillouin frequency of $9.2\,\text{GHz}$
equivalent to a sensitivity of $0.22\,\%/\text{nm}$ in excellent agreement with
Fig.~\ref{fig:example_fsbs}.
Still, Fig.~\ref{fig:example_fsbs} demonstrates that the geometry sensitivity of 
FSBS is only a response to size variations of a transverse acoustic resonator.

\begin{figure}
  \centering
  \includegraphics[width=0.7\textwidth]{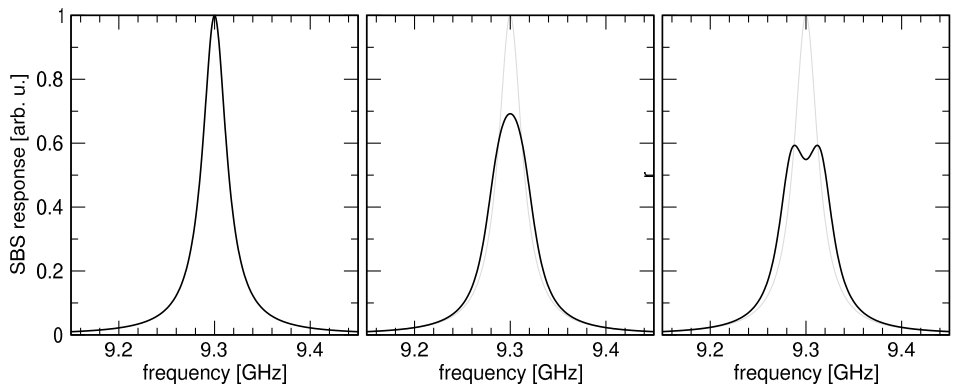}
  \caption{
    Effect of geometry-induced broadening on an SBS resonance.
    The system is forward SBS in a suspended silicon nanowire of dimensions
    $220\times450\,\text{nm}$ at a mechanical quality factor of $300$, \ie 
    the system studied in Fig~.\ref{fig:example_fsbs} and in 
    Ref.~\cite{vanLaer2015}.
    The three panels show the unperturbed resonance (left panel),
    the line for the case that the width linearly varies from $449$ to 
    $451\,\text{nm}$ over the waveguide length (center panel) and for
    the case that the waveguide width is sinusoidally undulated between
    the values $449$ and $451\,\text{nm}$.
    In both cases, the perturbation reduces the quality factor by about a 
    factor of two.
  }
  \label{fig:num_resonance_shapes}
\end{figure}

In addition, we have also computed broadened FSBS spectra of the first mode
(computed frequency shift: $9.3\,\text{GHz}$ at a width of $450\,\text{nm}$).
Fig.~\ref{fig:num_resonance_shapes} shows the impact of a width variation of 
$\pm1\,\text{nm}$.
This is a little less than two lattice constants of the silicon crystal, \ie 
it corresponds roughly to two monolayers of atoms.
Fig.~\ref{fig:num_resonance_shapes} also illustrates the effect of different 
distributions of width variations along the waveguide.
The leftmost panel shows the unperturbed spectrum assuming a mechanical quality 
factor of $300$, which is a realistic value for certain FSBS 
experiments~\cite{vanLaer2015} where mechanical loss is dominated by acoustic 
leakage.
The center panel shows the broadened spectrum assuming that the waveguide 
length linearly changes along the total waveguide length (typically several 
millimetres to centimetres); as a result the quality factor significantly 
drops to $183$.
The rightmost panel finally shows the broadened spectrum assuming that the
waveguide width fluctuates sinusoidally along the waveguide with a period that
is large compared to the acoustic decay length and the optical wave length 
(the exact periodicity does not matter).
In this case, not only is the quality factor further reduced to $155$, but the
resonance is also deformed into a highly non-Lorentzian shape with two distinct 
peaks.
This example shows that high effective quality factors in integrated SBS require
very consistent fabrication down to the atom level.

For BSBS, the situation is slightly more complicated, because the indirect 
(optical) perturbation contributes to the total sensitivity.
Therefore, we additionally determined the impact of the direct contribution by 
solving the acoustic problem for the perturbed geometry and the wave number 
derived from the unperturbed problem (circles in Fig.~\ref{fig:example_bsbs}); 
the indirect (optical) contribution (squares in Fig.~\ref{fig:example_bsbs}) 
then follows as the difference.
Both studied acoustic modes are hybrid and have an axial wave length that is
comparable to the waveguide's dimensions.
Therefore, the acoustic frequency is roughly determined by the bulk speed of 
sound and the wave number.
The displacement field of the first acoustic mode has appreciable 
$x$-components, which leads to a noticeable direct (acoustic) sensitivity.
In contrast, the second mode is mostly insensitive to the direct perturbation
effect.
Both acoustic modes are very susceptible to the indirect (optical) perturbation 
effect.
For example, the second acoustic mode responds to a width variation of 
$1\,\text{nm}$ with a resonance shift of up to $0.4\%$.
Correspondingly, a waveguide with a width of $300\,\text{nm}$ would have to be 
manufactured to better than $0.25\,\text{nm}$ (less than a monolayer of atoms) 
along its entire length in order to maintain an overall quality factor of $1000$.
Evidence for this problem has been observed experimentally~\cite{vanLaer2015c}.

One interesting detail of Fig.~\ref{fig:example_bsbs} is the fact that the 
direct and indirect contributions to the inhomogeneous broadening can be of 
comparable magnitude.
Furthermore, they will tend to have opposite signs, because an increase in 
waveguide width tunes the transverse acoustic resonator to lower frequencies 
via the direct (acoustic) perturbation mechanism, while an increase in the 
width also leads to higher optical mode indices, corresponding to greater 
wave numbers and eventually higher acoustic frequencies via the indirect 
(optical) perturbation mechanism.
This suggests the very promising idea to design a waveguide structure with a
deliberately \emph{strong} direct sensitivity in order to cancel the inevitable 
indirect sensitivity.
The result would be a waveguide design with greatly relaxed fabrication 
tolerances with respect to the waveguide width.
The general applicability of this concept to minimise the sensitivity with 
respect to one design parameter (in this case the width) is proven by the 
behavior of mode 1 in Fig.~\ref{fig:example_bsbs}.
There, the direct and indirect contributions for width variations nearly cancel 
for small waveguide widths.
However, the presented mode is only of academic interest because of its very
low SBS power gain.
Finally, it should be noted that only one specific type of perturbations (e.g. 
width variations, wall angle variations or height variations) can be 
compensated with this technique while potentially increasing the sensitivity
of the other perturbation types.

%%%%%%%%%%%%%%%%%%%%%%%%%%%%%%%%%%%%%%%%%%%%%%%%%%%%%%%%%%%%%%%%%%%%%%%%%%%%%%%%
%%%%%%%%%%%%%%%%%%%%%%%%%%%%%%%%%%%%%%%%%%%%%%%%%%%%%%%%%%%%%%%%%%%%%%%%%%%%%%%%
\section{Summary}
\label{sec:summary}

In conclusion, we have analysed the sensitivity of the optical and acoustic
dispersion relations on structural variations of a waveguide and related this
to the sensitivity of the Stokes shift $\Omega$.
We have shown that variations of the optical dispersion relation are a major
source of inhomogeneous broadening in backward SBS and that they are 
intimately related to the forward SBS coupling, because slowly and smoothly
varying perturbations mainly modify the phase of an optical mode, and induce
negligible back-scattering.
This leads to a counter-intuitive result in SBS dominated by radiation pressure 
(see \secref{sec:rad_BSBS}):
Since the inhomogeneous broadening is proportional to the forward SBS coupling
(\ie square root of FSBS-gain), the maximum of a strongly broadened SBS spectrum 
is proportional to the ratio of the backward gain and the square root of the 
forward gain.
As a result the maximum of the strongly broadened response of a backward
SBS system scales only with the square root of the naively expected SBS-gain.

This means that high-gain backward SBS waveguides are intrinsically sensitive 
to fabricational imperfections and require fabricational tolerances matched not
only to the intrinsic acoustic line width but also to the opto-acoustic coupling
strength.
Therefore, it may be advisable in practice to sacrifice some intrinsic SBS-gain
to avoid the tolerance-dominated regime, especially since this might reduce the
linear or nonlinear optical loss at the same time and thereby increase the 
overall figure of merit.
We have furthermore shown that the indirect resonance sensitivity is absent 
in forward SBS and that therefore no fundamental link between the resonance 
sensitivity and the SBS-gain exists in this case.
Nonetheless, the direct (acoustic) resonance sensitivity can be appreciable
and has been observed in at least one nanophotonic realisation of forward 
SBS~\cite{vanLaer2015}.
Our numerical example for FSBS is in excellent agreement with this experiment.
Finally, we have argued that the direct (acoustic) and indirect (optical)
contributions to the resonance sensitivity tend to have opposite sign.
This allows one to design comparatively insensitive high-gain waveguides with a 
strong direct sensitivity carefully engineered to compensate the intrinsic 
indirect sensitivity.

%%%%%%%%%%%%%%%%%%%%%%%%%%%%%%%%%%%%%%%%%%%%%%%%%%%%%%%%%%%%%%%%%%%%%%%%%%%%%%%%
%%%%%%%%%%%%%%%%%%%%%%%%%%%%%%%%%%%%%%%%%%%%%%%%%%%%%%%%%%%%%%%%%%%%%%%%%%%%%%%%
\section*{Acknowledgements}
C.W., M.J.S., B.J.E. and C.G.P. acknowledge financial support from the 
Australian Research Council (ARC) via the Discovery Grant DP130100832, 
its Laureate Fellowship (B.J.E., FL120100029) program,
and the ARC Center of Excellence CUDOS (CE110001018).
R.V.L. acknowledges the Agency for Innovation by Science and Technology in 
Flanders (IWT) for a PhD grant.

%%%%%%%%%%%%%%%%%%%%%%%%%%%%%%%%%%%%%%%%%%%%%%%%%%%%%%%%%%%%%%%%%%%%%%%%%%%%%%%%
%%%%%%%%%%%%%%%%%%%%%%%%%%%%%%%%%%%%%%%%%%%%%%%%%%%%%%%%%%%%%%%%%%%%%%%%%%%%%%%%
\appendix
\section{Mechanical edge perturbations}
\label{appx:mech_edge}

In this appendix, we derive the edge contribution to the mechanical perturbation
term.
This is analogous to the perturbation theory developed by Johnson et al. for 
the perturbation of the electromagnetic dispersion relation~\cite{Johnson2002}, 
which is based on the continuity of the transversal electric field and of the 
normal electric induction field across a material boundary.
In the context of continuum mechanics, the quantities that are continuous across
a material boundary are the normal projection of the stress tensor 
\begin{align}
  T^\perp_{ij} = & n_i \sum_k n_k T_{kj},
\end{align}
and the in-plane projection of the strain tensor
\begin{align}
  S^\parallel_{ij} = & S_{ij} - n_i \sum_k n_k S_{kj},
\end{align}
where $\hat n$ is the surface normal.
In analogy to the electromagnetic case, the perturbation to the Hamiltonian
is the normal projection of the surface displacement multiplied by the 
mechanical energy density expressed in terms of the conserved quantities,
where we restrict ourselves to isotropic materials such as glasses or 
homogenised polycrystalline cubic materials (the expressions for anisotropic
materials are more complex, but of the same general form).
We introduce an auxiliary strain quantity 
\begin{align}
  S^\eff_{ij} = S^\parallel_{ij} + n_i \sum_{klm} n_k [c^{-1}]_{kjlm} T^\perp_{lm},
  \label{eqn:s_eff}
\end{align}
where $[c^{-1}]$ is the mechanical compliance tensor.  
This allows us to write the acoustic perturbation overlap caused by 
boundary displacements in a concise form:
\begin{align}
  \perturb^{\ac,\edge} = \int_\edge \total \myvec r \ 
  \sum_{ijkl} \left[S^\eff_{ij} c_{ijkl} S^\eff_{kl} \right] (\hat n \cdot \myvec U).
\end{align}
Within the modal expansion employed in \secref{sec:theory}, the edge-effect 
of the $p$-th acoustic basis function of the phase matched acoustic mode $n_b$ 
becomes:
\begin{align}
  R^\edge_p = & \int_\edge \total \myvec r \ 
  \sum_{ijkl} \left[\left(s^\eff_{ij}\right)^\ast c_{ijkl} s^\eff_{kl} \right] 
  (\hat n \cdot \myvec u^{(p)})^\ast,
  \intertext{with the modal (note the lower case) generalisation of \eqnref{eqn:s_eff}}
  s^\eff_{ij} = &
  \frac{1}{2} \Bigg[
  \partial_i u^{(n_b)}_j + \partial_j u^{(n_b)}_i
  - n_i \sum_k n_k \Big( \partial_k u^{(n_b)}_j + \partial_j u^{(n_b)}_k \Big)
  \\
  \nonumber & \quad
  + n_i \sum_{klpq\alpha\beta} n_k 
  [c^{-1}]_{kjlp} n_l n_q c_{pb\alpha\beta} 
  \Big(\partial_\alpha u^{(n_b)}_\beta + \partial_\beta u^{(n_b)}_\alpha \Big)
  \Bigg].
\end{align}

%%%%%%%%%%%%%%%%%%%%%%%%%%%%%%%%%%%%%%%%%%%%%%%%%%%%%%%%%%%%%%%%%%%%%%%%%%%%%%%%
\section{Ratio of FSBS and BSBS radiation pressure terms}
\label{appx:gain_ratio}

In this appendix we investigate the ratio between the acousto-optic coupling 
terms of forward and backward SBS.
We find that they are always of the same order and that the radiation pressure
coupling is always greater for forward SBS.

Inserting $\myvec e^{(1)} = [\myvec e^{(2)}]^\ast = \myvec e$ into the explicit 
definitions Eqs.~(\ref{eqn:def_q_bulk},\ref{eqn:def_q_edge}) the forward and 
backward coupling terms are:
\begin{align}
  Q^\FSBS_p = & Q^{\bulk,\FSBS}_{p} + Q^{\edge,\FSBS}_{p}
  = Q^\bulk_{11;p} + Q^\edge_{11;p}
  \\
  = & 
  \eps_0 \int_\bulk \total^2 r \ \sum_{klmn}
  \eps_r^2 e_k^\ast e_l p_{klmn} \partial_m [u^{(p)}_n]^\ast 
  \\
  \nonumber & \quad
  +
  \int_{\edge} \total \myvec r \ 
  \Big[
    (\eps_a - \eps_b) \eps_0 (\hat n \times \myvec e^\ast (\hat n \times \myvec e)
    -
    (\eps^{-1}_a - \eps^{-1}_b) \eps^{-1}_0 (\hat n \cdot \myvec d)^\ast (\hat n \cdot \myvec d)
  \Big]
  (\hat n \cdot \myvec u^{(p)})^\ast ,
  \\
  Q^\BSBS_p = & Q^{\bulk,\BSBS}_{p} + Q^{\edge,\BSBS}_{p}
  = Q^\bulk_{12;p} + Q^\edge_{12;p}
  \\
  = & 
  \eps_0 \int_\bulk \total^2 r \ \sum_{klmn}
  \eps_r^2 e_k^\ast e_l^\ast p_{klmn} \partial_m [u^{(p)}_n]^\ast 
  \\
  \nonumber & \quad
  +
  \int_{\edge} \total \myvec r \ 
  \Big[
    (\eps_a - \eps_b) \eps_0 (\hat n \times \myvec e^\ast (\hat n \times \myvec e^\ast)
    -
    (\eps^{-1}_a - \eps^{-1}_b) \eps^{-1}_0 (\hat n \cdot \myvec d)^\ast 
    (\hat n \cdot \myvec d^\ast)
  \Big]
  (\hat n \cdot \myvec u^{(p)})^\ast .
\end{align}
The only difference is the complex conjugation of the second optical mode.
Due to the symmetry of Eqs.~(\ref{eqn:ewp_opt},\ref{eqn:ewp_ac}) the phases 
of the optical as well as the acoustic modes can be adjusted such that the 
in-plane field components are purely real-valued and the axial components are 
purely imaginary.
Thus, the main effect of the complex conjugation is a sign reversal in the
$z$-component of the second optical mode.
In the photoelastic coupling term $Q^\bulk_{ij;n}$ the axial components of 
one mode and the transversal components of another mode can be nontrivially 
combined depending on the orientation of the principal axes of $p_{ijkl}$ 
relative to the waveguide geometry.
However, the edge term $Q^\edge_{ij;n}$ can be easily analysed.
First we notice that the axial component of $\myvec u^{(p)}$ is irrelevant.
Next, we decompose the transversal electromagnetic fields into components 
normal and parallel to the waveguide surface:
\begin{align}
  e_\perp = & \hat n \cdot \myvec e;
  \\
  e_\parallel = & 
  |\myvec e - \hat n (\hat n \cdot \myvec e) - \hat z (\hat z \cdot \myvec e)|.
\end{align}
Within this notation we find:
\begin{align}
  Q^{\edge,\FSBS}_p = & 
  \int_{\edge} \total \myvec r \ 
  \Big[
    (\eps_a - \eps_b) \eps_0 (|e_\parallel|^2 + |e_z|^2)
    +
    (\eps^{-1}_b - \eps^{-1}_a) \eps^{-1}_0 |d_\perp|^2 
  \Big] (\hat n \cdot \myvec u^{(p)})^\ast,
  \\
  Q^{\edge,\BSBS}_p = & 
  \int_{\edge} \total \myvec r \ 
  \Big[
    (\eps_a - \eps_b) \eps_0 (|e_\parallel|^2 - |e_z|^2)
    +
    (\eps^{-1}_b - \eps^{-1}_a) \eps^{-1}_0 |d_\perp|^2 
  \Big] (\hat n \cdot \myvec u^{(p)})^\ast.
\end{align}
The factors $(\eps_a - \eps_b)$ and $(\eps_b^{-1} - \eps_a^{-1})$ always have
the same sign (note the reversal of subscripts).
Therefore, the radiation pressure contribution to the intra-mode forward SBS 
coupling is always greater than the corresponding contribution to the backward 
SBS coupling.
This result does not apply to the photoelastic coupling.

%\TODO{Appendix on Poisson brackets?}

%\bibliography{bibliography,extra}

\begin{thebibliography}{99}


\bibitem{Boyd2003}
  R.~W. Boyd,
  \newblock {\em {Nonlinear optics}}
  \newblock (Academic, 3rd ed., 2003).

\bibitem{Agrawal}
  G. P. Agrawal,
  \newblock {\em {Nonlinear fiber optics}}
  \newblock (Academic, 5th ed., 2012).

\bibitem{Brillouin1922}
  L. Brillouin.
  ``Diffusion de la lumi\`{e}re par un corps transparent homog\`{e}ne,''
  Annals of Physics \textbf{17,} 88--122 (1922).

\bibitem{Chiao1964}
  R.~Y. Chiao, C.~H. Townes, and B.~P Stoicheff,
  ``Stimulated Brillouin scattering and coherent generation of intense hypersonic waves,''
  Phys. Rev. Lett. \textbf{12,} 592 (1964).

\bibitem{Uchida1973}
  N. Uchida and N. Niizeki.
  ``Acoustooptic Deflection Materials and Techniques,''
  Proc. IEEE \textbf{61,} 1073--1092 (1973).

\bibitem{Dainese2006}
  P.~Dainese, P.~St.~J. Russell, N.~Joly, J.~C. Knight, G.~S. Wiederhecker, 
  H.~L. Fragnito, V.~Laude, and A.~Khelif,
  ``Stimulated Brillouin scattering from multi-GHz-guided acoustic phonons 
  in nanostructured photonic crystal fibres,''
  Nature Phys. \textbf{2,} 388--392 (2006).

\bibitem{Eggleton2013}
  B.~J. Eggleton, C.~G. Poulton, and R.~Pant,
  ``Inducing and harnessing stimulated Brillouin scattering in photonic integrated circuits,''
  Adv. Opt. Photonics \textbf{5,} 536--587 (2013).

\bibitem{Pant2011}
  R. Pant, C.~G. Poulton, D.-Y. Choi, H.~Mcfarlane, S.~Hile, E.~Li, L.~Th{\'e}venaz, B.~Luther-Davies, S.~J. Madden, and B.~J. Eggleton,
  ``On-chip stimulated Brillouin scattering,''
  Opt. Express \textbf{19,} 8285--8290 (2011).

\bibitem{Rakich2012}
  P.~T. Rakich, C.~Reinke, R.~Camacho, P.~Davids, and Z.~Wang,
  ``Giant enhancement of stimulated Brillouin scattering in the subwavelength Limit,''
  Phys. Rev. X \textbf{2,} 011008 (2012).

\bibitem{Shin2013}
  H.~Shin, W.~Qiu, R.~Jarecki, J.~A. Cox, R.~H. O.~Ill, A.~Starbuck, Z.~Wang, and P.~T. Rakich,
  ``Tailorable stimulated Brillouin scattering in nanoscale silicon waveguides,''
  Nature Commun. \textbf{4,} 1944 (2013).

\bibitem{vanLaer2015}
  R.~Van Laer, B.~Kuyken, D.~Van Thourhout, and R.~Baets,
  ``Interaction between light and highly confined hypersound in a silicon
  photonic nanowire,'' Nature Photon. \textbf{9,} 199–-203 (2015).

\bibitem{Chin2010}
  S. Chin, L. Th\'{e}venaz, J. Sancho, S. Sales, J. Capmany, P. Berger, J. Bourderionnet, and D. Dolfi,
  ``Broadband true time delay for microwave signal processing, using slow light based on stimulated Brillouin scattering in optical fibers,''
  Opt. Express \textbf{18,} 22599--22613 (2010).

\bibitem{Abedin2012}
  K.~S. Abedin, P.~S. Westbrook, J.~W. Nicholson, J. Porque, T.  Kremp, and X. Liu,
  ``Single-frequency Brillouin distributed feedback fiber laser,''
  Opt. Lett. \textbf{37,} 605 (2012).

\bibitem{Grudinin2010}
  I.~S. Grudinin, H. Lee, O. Painter, and K.~J. Vahala,
  ``Phonon Laser Action in a Tunable Two-Level System,''
  Phys. Rev. Lett. \textbf{104,} 083901 (2010).

\bibitem{Kovalev2002} 
  V.~I. Kovalev, and R.~G. Harrison, 
  ``Waveguide-induced inhomogeneous spectral broadening of stimulated 
  Brillouin scattering in optical fiber, ''
  Opt. Lett. \textbf{27,} 2022--2024 (2002).

\bibitem{Wolff2015b}
  C. Wolff, M.~J. Steel, B.~J. Eggleton, and C.~G. Poulton,
  ``Acoustic build-up in on-chip stimulated Brillouin scattering, ''
  Sci. Rep. \textbf{5,} 13656 (2015).

\bibitem{Beugnot2011}
  J.-C. Beugnot, M.~Tur, S.~Foaleng Mafang, and L.~Th{\'e}venaz, 
  ``Distributed Brillouin sensing with sub-meter spatial resolution: modeling and processing,'' 
  Opt. Express \textbf{19,} 7381--7397 (2011).

\bibitem{Stiller2012}
  B.~Stiller, A.~Kudlinski, Min Won Lee, G.~Bouwmans, M.~Delqu{\'e}, 
  J.-C.~Beugnot, H.~Maillotte, and T.~Sylvestre, 
  ``SBS Mitigation in a Microstructured Optical Fiber by Periodically Varying 
  the Core Diameter, ''
  IEEE Photon. Technol. Lett. \textbf{24,} 667--669 (2012).

\bibitem{vanLaer2015b}
  R.~Van Laer, B.~Kuyken, R.~Baets, D.~Van Thourhout
  ``Unifying Brillouin scattering and cavity optomechanics, ''
  {\it arXiv:1503.03044 [physics.optics]}, (2015).

\bibitem{Aspelmeyer2014}
  M.~Aspelmeyer, T.~J. Kippenberg, and F.~Marquardt, 
  ``Cavity optomechanics, ''
  Rev. Mod. Phys.  \textbf{86,} 1391--1452 (2014).

\bibitem{Sipe2015}
  J.~Sipe, and M.~J. Steel
  {``A Hamiltonian treatment of stimulated Brillouin scattering in nanoscale 
  integrated waveguides,''}
  {\it arXiv:1509.01017 [physics.optics]}, (2015).

\bibitem{Wolff2015a}
  C.~Wolff, M.~J. Steel, B.~J. Eggleton, and C.~G. Poulton,
  {``Stimulated Brillouin Scattering in integrated photonic waveguides: forces,
  scattering mechanisms and coupled mode analysis,''}
  Phys. Rev. A \textbf{92}, 013836 (2015).

\bibitem{Horiguchi1995}
  T.~Horiguchi, K.~Shimizu, T.~Kurashima, M.~Tateda, Y.~Koyamada, 
  ``Development of a distributed sensing technique using Brillouin scattering,''
  J. Lightwave Technol., \textbf{13}, 1296--1302 (1995).

\bibitem{Nikles1996}
  M.~Nikles, L.~Thevenaz, P.~A. Robert,
  ``Simple distributed fiber sensor based on Brillouin gain spectrum analysis,''
  Opt. Lett. \textbf{21}, 758-760 (1996).

\bibitem{Bigoni}
  D.~Bigoni
  \newblock {\em {Nonlinear Solid Mechanics}}
  \newblock (Cambridge, 2012).

\bibitem{Beugnot2012}
  J.-C. Beugnot, V.~Laude, 
  ``Electrostriction and guidance of acoustic phonons in optical fibers,''
  Phys. Rev. B \textbf{86}, 224304 (2012).

\bibitem{Johnson2002}
  S.~G. Johnson, M. Ibanescu, M.~A. Skorobagotiy, O. Weisberg, J.~D. Joannopoulos, Y. Fink,
  ``Perturbation theory for Maxwell's equations with shifting material boundaries,''
  Phys. Rev. E \textbf{65,} 066611 (2002).

\bibitem{vanLaer2014}
  R. Van~Laer, B.~Kuyken, D. Van~Thourhout, and R.~Baets, 
  ``Analysis of enhanced stimulated Brillouin scattering in silicon slot waveguides,''
  Opt. Lett. \textbf{39,} 1242-1245 (2014).

\bibitem{Wolff2014}
   C.~Wolff, M.~J. Steel, C.~G. Poulton,
   ``Formal selection rules for Brillouin scattering in integrated waveguides and structured fibers,''
   Opt. Express \textbf{22}, 32489-32501 (2014).

\bibitem{vanLaer2015c}
  R.~Van~Laer, A.~Bazin, B.~Kuyken, R.~Baets, D.~Van~Thourhout,
  ``Net on-chip Brillouin gain based on suspended silicon nanowires,''
  New. J. Phys. \textbf{17}, 115005 (2015).


\end{thebibliography}
%\bibliographystyle{osajnl}

\end{document}